\documentclass{aa}
\usepackage[dvips]{graphicx}
\newcommand{\asec}{$^{\prime\prime}$}
\newcommand{\pas}{.\hskip-2pt$^{\prime\prime}$}
\def\I{IRAS\,23385+6053}
\def\II{I23385}
\def\AMM{NH$_3$}
\def\ace{CH$_{3}$C$_{2}$H}
\def\CI{$^{13}$CO}
\def\CII{C$^{18}$O}
\def\CIII{C$^{17}$O}

\def\HCOp{\mbox{HCO$^+$}}

\def\HII{H{\sc ii}}

\def\kms{\mbox{km~s$^{-1}$}}
\def\cmc{cm$^{-3}$}
\def\cmq{cm$^{-2}$}

\def\Tk{\mbox{$T_{\rm k}$}}

\def\mvir{\mbox{$M_{\rm VIR}$}}
\begin{document}
\title{IRAS 23385+6053: a candidate protostellar massive object
\thanks{Based on observations carried out with the IRAM Plateau de Bure
Interferometer.
IRAM is supported by INSU/CNRS (France), MPG (Germany) and IGN (Spain).}}
\author{F. Fontani \inst{1} \and R. Cesaroni \inst{2}
	\and L. Testi \inst{2} \and C.M. Walmsley \inst {2}
	\and S. Molinari \inst{3} 
        \and R. Neri \inst{4} \and D. Shepherd \inst{5}  
        \and J. Brand \inst {6} \and F. Palla \inst{2} \and Q. Zhang \inst{7}}
\institute{Dipartimento di Astronomia e Fisica dello spazio, Largo E. Fermi 2,
           I-50125 Firenze, Italy \and
	   INAF, Osservatorio Astrofisico di Arcetri, Largo E. Fermi 5,
           I-50125 Firenze, Italy \and
           IFSI, CNR, Via Fosso del Cavaliere, I-00133 Roma, Italy\and
           Institut de Radio Astronomie Milimetrique, 300 Rue de la Piscine, 
           F-38406 St. Martin de Heres, France  \and
           National Radio Astronomy Observatory, P.O. Box O, Socorro, NM 87801 
           \and
	   Istituto di Radioastronomia, CNR, Via Gobetti 101, 40129 Bologna,
	   Italy \and
           Harvard Smithsonian Center for Astrophysics, 60 Garden Street,
           Cambridge, MA 02138
 }
\offprints{F. Fontani, \email{fontani@arcetri.astro.it}}
\date{Received date; accepted date}

\titlerunning{IRAS 23385+6053}
\authorrunning{Fontani et al.}

\abstract{We present the results of a multi-line and continuum study 
towards the source \I\
 performed with the IRAM-30m telescope, the Plateau de Bure 
Interferometer, the Very Large Array Interferometer and the James Clerk 
Maxwell Telescope. We have obtained single-dish maps in the \CII\ (1--0),
 \CIII\ (1--0) and (2--1) rotational lines, interferometric maps in the 
\ace\ (13--12) line, \AMM\ (1,1) and (2,2) inversion transitions, and single-pointing observations of the \ace\ (6--5), (8--7) and (13--12) rotational
lines. 
The new results confirm our earlier findings, namely that \I\ is a good candidate high-mass protostellar object, precursor 
of an ultracompact \HII\ region.
The source is roughly composed of two regions: a
molecular core $\sim0.03\div0.04$ pc in size, with a temperature of 
$\sim40$ K and
an H$_{2}$ volume density of the order of 10$^{7}$ \cmc, and an extended halo
of diameter $\leq$0.4~pc, with an average kinetic temperature of
$\sim 15$ K and H$_{2}$ volume density of the order of 10$^{5}$ \cmc. The
core temperature is much smaller than what is typically found in molecular
 cores of the same diameter surrounding massive ZAMS stars. 
From the continuum spectrum we deduce that the core luminosity is
between 150 and $1.6\times10^{4}L_{\odot}$, and we believe that the upper 
limit is near the ``true'' source luminosity.
Moreover, by comparing the H$_{2}$ volume density
obtained at different radii from the IRAS source, we find that the halo
has a density profile of the type
$n_{\rm H_{2}}\propto r^{-2.3}$. This suggests that the source is 
gravitationally unstable. The latter hypothesis is also supported by 
a low virial-to-gas mass ratio ($M_{\rm VIR}/M_{\rm gas}$ $\leq$0.3).
Finally, we demonstrate that the temperature at the core surface
is consistent with a core luminosity of $10^3\;L_{\odot}$ and conclude
that we might be observing a protostar still accreting material
from its parental cloud, whose mass at present is $\sim 6 M_{\odot}$.
}

\maketitle

\section{Introduction}
\label{intro}
The study of massive stars ($M\geq 10~M_\odot$) and their formation
is important for a better understanding of the evolution and 
morphology of the Galaxy. However, up to now most progress has been done in the
study of the formation of low-mass 
stars ($M\leq 1~M_\odot$). This is a consequence of the many observational 
problems which
hinder the study of the high-mass star formation process: massive stars 
 are more distant than low-mass ones, interact more strongly with their
 environment, have shorter evolutionary timescales and mainly form in
clusters. Neverthless, recently a major observational
effort has been made to identify the very earliest stages of their 
evolution. Successful results have been obtained searching for high-density and
high-temperature tracers (e.g. NH$_3$ or CH$_3$CN) towards selected
targets associated with regions of massive star formation, such as
ultracompact \HII\ regions (Cesaroni et al. \cite{cesa94}; Olmi et al. \cite{
olmi96}; Cesaroni et al. \cite{cesa98}),
H$_2$O masers (Codella et al. \cite{code}; Plume et al. \cite{plume}; Beuther et al. \cite{beuther2}),
and IRAS sources (Molinari et al. \cite{mol96}; Beuther et al. \cite{beuther}).
Probably the most relevant finding of these studies is the detection of
hot ($\ga$100~K), dense ($\ga 10^7$~\cmc), molecular cores where high-mass
 stars have recently formed. It is believed that such ``hot cores'' 
(HCs, hereafter) represent the natal environment of O--B stars.

The next step in the search for very young massive stars is the detection
of a genuine example of massive protostar. 
Molinari et al. (\cite{mol98b}) suggested
that the source \I\ might be such an object.

\I\ is one of a sample of IRAS sources selected in a long-standing
project as a likely candidate massive protostar. 
 The selection criteria used in this project
are based on the IRAS colours (Palla et al. \cite{palla91}) and are aimed at 
identifying luminous
(and hence likely massive) stellar objects in the earliest stages of their
evolution. Observations of molecular tracers such as H$_2$O
(Palla et al. \cite{palla91}) and NH$_3$
 (Molinari et al. \cite{mol96})
have proven these objects to be associated with dense molecular
gas. Additional observations in the continuum with the James Clerk Maxwell
Telescope (JCMT, Molinari et al. \cite{mol00}) and the Very Large Array (VLA) 
(Molinari et al. \cite{mol98a})
have confirmed that a well defined sub-sample of these sources is
embedded in dense dusty clumps (detected at mm and sub-mm wavelengths),
and do not present any free-free emission, i.e. are undetected
at centimeter wavelengths down to levels below that expected on the basis
of their luminosity.
The latter finding suggests that, although luminous (i.e. massive), the
objects embedded in the clumps are still too young to develop an \HII\ region.
Such a sample contains excellent targets to search for massive
protostars.

Recently, Brand et al. (\cite{brand}) have performed a single-dish multi-line
 study mapping a small number of these candidates in various molecular 
transitions,
finding that they are associated with clumps that are larger, cooler, more 
massive and less turbulent than those associated with ultracompact 
\HII\ regions.

One of these sources, \I\ (at a kinematic distance of $\sim 4.9$ kpc), was 
observed in more detail at different angular resolutions
with the OVRO interferometer (Molinari et al.~\cite{mol98b}, \cite{mol02})
and the VLA (Molinari et al.~\cite{mol02}).
This object has also been observed with ISOCAM at 7 and 15~$\mu$m.
The main findings of these first studies are the following: 
\begin{itemize}
\item the source presents a compact molecular, dusty core with diameter 
$\sim$0.08~pc and mass $\sim$370~$M_\odot$, centered at 
$\alpha({\rm J2000})={\rm 23^{h}40^{m}}$54\fs 5, 
$\delta({\rm J2000})=61^{\circ}10'$28\pas 10. 
{\it Hereafter, the name \II\ will be used to identify this core};
\item the luminosity
 derived from the IRAS and JCMT continuum measurements for a distance of 
4.9~kpc is 
$\sim$$1.6~10^4~L_\odot$;
\item {\it no} continuum emission from the core is 
detected at 2 and 6~cm with the VLA, down to $\sim$0.5~mJy beam$^{-1}$ (3$\sigma$);
\item {\it no} continuum emission from the core is detected at 7 and 15~$\mu$m
 with ISOCAM 
 down to $\sim$6~mJy beam$^{-1}$ (3$\sigma$); 
\item a faint, small ($\le$0.1~pc) bipolar outflow with axis closely aligned
with the line of sight has been seen in the SiO, HCO$^+$, \CI\ and CS line 
wings.
\end{itemize}
Molinari et al. (\cite{mol98b}) concluded that \II\ is a good example of a
luminous,
young, deeply embedded (proto)stellar source. 

Also, Molinari et al.~(\cite{mol02}) discovered two 
extended \HII\ regions with the VLA at 3.6~cm which, however, do not
coincide with \II ; rather they seem to overlap with a cluster of 
infrared sources detected with ISOCAM that surrounds the molecular peak.
The infrared emission is coincident with a cluster of stars in 
near-infrared images (Molinari, priv. comm.).
In this work, we 
discuss whether \II\ is a newly-born B0 star just arrived on the ZAMS
or if it is a massive protostar still in the main accretion phase.
For this purpose, the question
we must answer is whether \II\ derives its luminosity primarly
from hydrogen burning or from accretion.
We believe that understanding the nature of \II\ can be 
improved by examining the physical parameters of the core, in particular
by measuring the {\it temperature} of the associated core. 
In fact, in low-mass objects the temperature seems to be lower in
the earlier evolutionary stages (Myers \& Ladd 1993). By analogy, if 
high-mass ZAMS stars are found in
``hot cores'', then high-mass {\it proto}stars might
lie inside colder cores.
Furthermore, a detailed picture of the source, from
small to large spatial scales, should be very useful to our purposes.
We present here the results obtained from observations of the \CII\ and
 \CIII\ (1--0), the \CIII\ (2--1) and the \ace\ (6--5), (8--7) and (13--12) 
lines
using the IRAM-30m telescope and the Plateau de Bure Interferometer (PdBI).
We also mapped the \AMM\ (1,1) and (2,2) inversion lines with the VLA.
In Sect. 2 we will describe the observations; in Sects. 3 and 4 we will present
the observational results and derive the physical parameters of the source,
respectively;
in Sect. 5 we will discuss these results, and finally we will draw our 
conclusions in Sect. 6. 

\section{Observations and data reduction}

\subsection{IRAM-30m}

The molecular transitions observed with the IRAM-30m telescope
are listed in Table 1: here, we also give the frequencies, the half power
beam width of the telescope, the total bandwidth used and the spectral
resolution. The main beam brightness temperature,
$T_{\rm MB}$, and the flux density, $F_{\nu}$, are related by the expression
$F_{\nu}({\rm Jy})=4.9\;T_{\rm MB}(\rm K)$.

\subsubsection{\CII\ and \CIII}

\CII\ and \CIII\ data were obtained on September 5, 2000. We simultaneously
used 
two 3~mm receivers centered at the frequencies of the \CII\ (1--0) and \CIII\
(1--0), and two 1.3~mm receiver both centered at the frequency of the \CIII\
(2--1) line to optimize the signal-to-noise ratio. For every
line, three 3$\times$3 maps with grid spacings 6, 8 and 10\asec\ and three 
5$\times$5 point maps with the same grid spacing were made. All maps are 
centered at
$\alpha({\rm J2000})={\rm 23^{h}40^{m}}$54\fs 5, 
$\delta({\rm J2000})=61^{\circ}10'$28\pas 10: this position corresponds to 
the sub-mm 
peak detected by Molinari et al. (\cite{mol00}). Map sampling allows us to 
cover a region $\sim40$\asec\ in size. Pointing and receiver
 alignment were regularly checked, and they were found to be
accurate to within 2\asec . The data were calibrated with the ``chopper wheel''
technique (see Kutner \& Ulich \cite{kutner}).
The integration time was 2 min per point in ``wobbler-switching'' 
mode, namely a nutating secondary reflector with a beam-throw
of 240\asec\ in azimuth and a phase duration of 2 s. 

The system temperature was $\sim250$ K at the frequency of the \CIII\ (1--0) 
and \CII\ (1--0) lines, and it was $\sim 1000$ K in the \CIII\ (2--1) line.

\subsubsection{\ace\ }

\ace\ data were obtained on August 12, 1999. We simultaneously 
observed the (6--5), (8--7) and (13--12) rotational transitions in the 
3~mm, 2~mm and 1.3~mm bands respectively. Only single-pointing observations
were made.
The data have been calibrated using the ``chopper wheel'' technique, and the
observations were performed using the ``wobbler-switching'' mode. 
The system temperature was $\sim120$ K at 3~mm, $\sim260$ K at 2~mm and 
$\sim350$ K at 1.3~mm.

\begin{table*}
\begin{center}
\caption{Transitions observed with the IRAM-30m telescope and the Plateau de
 Bure interferometer}
\label{30mobs}
\begin{tabular}{cccccc}
\hline \hline
 Molecular  &  Rest freq. & HPBW (30m)   & Band / Resolution (30m) & HPBW
(PdBI) & Band / Resolution (PdBI) \\
  transition    &    (GHz)    & (\asec )  & (MHz) & (\asec ) & (MHz) \\
\hline
\CII\ (1--0) &  109.782  & 23   &   70 / 0.078 ($\sim0.2$ \kms )  & 2.3$\times$1.94 & 16 / 0.156 ($\sim0.4$ \kms ) \\ 
\CIII\ (1--0) & 112.359  &  22   &   70 / 0.078 ($\sim0.2$ \kms) & 1.95$\times$1.45 & 17 / 0.156 ($\sim0.4$ \kms )
\\ 
\CIII\ (2--1) & 224.714   &  11  &   70 / 0.078 ($\sim0.1$ \kms ) & 0.94$\times$0.75 & 25 / 0.312 ($\sim0.4$ \kms )\\ 
\ace\ (6--5) & 102.547$(^*)$ &  24  &   70 / 0.078 ($\sim0.2$ \kms ) & -- & -- \\ 
\ace\ (8--7) & 136.728$(^*)$ &  18  &   70 / 0.078 ($\sim0.2$ \kms )& -- & -- \\
\ace\ (13--12) & 222.166$(^*)$ &  11  &   140 / 0.078 ($\sim0.1$ \kms ) & 0.94$\times$0.76 & 33 / 0.312 ($\sim0.4$ \kms ) \\ \hline
\end{tabular}
\end{center}
$(^*)$ rest frequency of the $K=0$ transition \\
\end{table*}

\begin{figure}
\centerline{\includegraphics[angle=0,width=8cm]{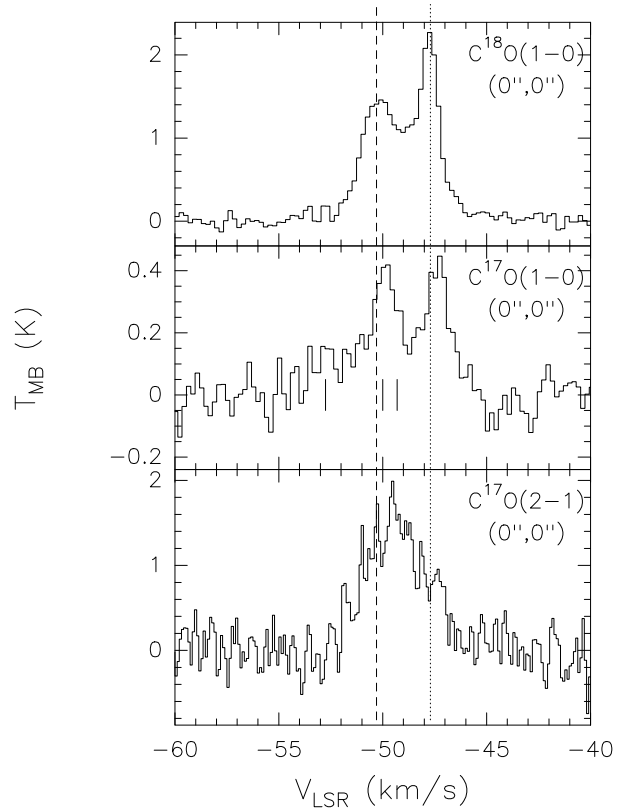}}
\caption{\CII\ and \CIII\ spectra towards the central position in the maps,
which is the peak of the sub-mm emission (Molinari et al. 2000). The dashed
line indicates the main component arising from the source.
 The dotted line with red-shifted velocity shows the second component. In the
\CIII\ (1--0), the vertical lines under the spectrum indicate the position of 
the hyperfine components.}
\label{fspec}
\end{figure}
\begin{figure}
\centerline{\includegraphics[angle=0,width=8cm]{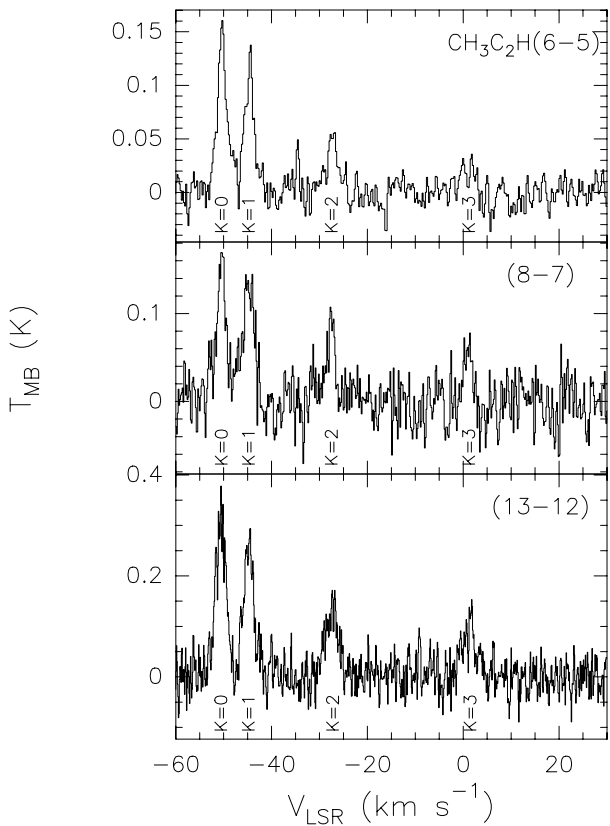}}
\caption{From top to bottom: \ace\ (6--5), (8--7) and (13--12) IRAM-30m 
spectra towards the peak of the sub-mm emission (Molinari et al. 2000). The 
numbers under the spectra indicate the position
of the different $K$ components. The $V_{\rm LSR}$ is relative to the line
with $K=0$.}
\end{figure}

\subsection{JCMT}

An 850$\mu$m continuum image was taken on October 16 1998 with SCUBA at the
JCMT (Holland et al. \cite{holland}) towards the postion of the core
\II . The standard 
64-points jiggle map observing mode was used, with a chop
throw of 2 arcmin in the SE direction. Atmospheric conditions were not
excellent, with $\tau_{225 GHz} \sim$ 0.15. Telescope focus and pointing
were checked using Uranus and the data were calibrated following standard 
recipes as in the SCUBA User Manual (SURF).

\subsection{Plateau de Bure Interferometer}

All transitions observed with the IRAM-30m telescope except for \ace\ (6--5) and
(8--7) have been observed also with the IRAM 5-element array at Plateau de
Bure. The observations were carried out in November 1998 and April 1999
in the B and C configurations of the array. The antennae were equipped with
82--116~GHz and 210--245~GHz receivers operating simultaneously. These were
tuned single side-band at 3~mm and double side-band at 1.3~mm resulting in
typical system temperatures of 200~K (in the USB) and 1000~K (in the LSB) at
3~mm and 250~K (DSB) at 1.3~mm. The facility correlator was centred at
112.581~GHz in the USB at 3~mm and at 221.921~GHz in the LSB at 1.3~mm. The
C$^{18}$O(1--0) and C$^{17}$O(1--0) lines were covered using two correlator
units of 20~MHz bandwidth (with 256 channels); two 160~MHz units were
suitably placed to obtain a continuum measurement at 3~mm. The
C$^{17}$O(2--1) and CH$_3$C$_2$H(13--12) $K$=0,1,2 transitions were observed
with a 40~MHz bandwidth (256 channels), while the other $K$ lines were
covered with 160~MHz bandwidths (64 channels), which were also used to obtain
a continuum measurement at 1.3~mm. The effective spectral resolution is about
twice the channel spacing (see ``The New Correlator Description of the PdBI'', http://www.iram.fr).

Phase and amplitude calibration were obtained by regular observations (every
20 min) of nearby point sources (2200+420 and 0059+5808). The bandpass
calibration was carried out on 3C454.3, while the absolute flux density scale
was derived from MWC349, regulary monitored at
IRAM. Continuum images were produced after averaging line-free channels and
then subtracted from the line. The resulting maps were then cleaned and
channel maps for the lines were produced. The synthesized beam sizes are 
listed in Table~\ref{30mobs}.

\subsection{Very Large Array}

The NRAO\footnote{The
National Radio Astronomy Observatory is a facility of the National Science
Foundation operated under cooperative agreement by Associated Universities,
INC.} Very Large Array (VLA) ammonia observations were performed on August
8, 2000. The NH$_3$(1,1) and NH$_3$(2,2) inversion lines at
23.694496 and 23.722633~GHz, respectively, were simultaneously observed 
with spectral resolutions of
1.2 and 4.9 \kms\ respectively. The array was used in its most
compact configuration (D), offering baselines from 35~m to 1~km.
The flux density scale was established by observing 3C286 and 3C147;
the uncertainty is expected to be less than 15\%. Gain calibration
was ensured by frequent observations of the point source J0102$+$584,
with a measured flux density at the time of the observations of 3.4~Jy.
The same calibrator was also used for bandpass correction.

The data were edited and calibrated using the Astronomical Image Processing
System (AIPS) following standard procedures.
Imaging and deconvolution was performed using the IMAGR task and 
naturally weighting the visibilities. The resulting synthesised beam
FWHM is 4.0$^{\prime\prime}\times 3.7^{\prime\prime}$ with position angle 
$-50^{\circ}$, and the noise level in each channel is 4~mJy beam$^{-1}$ and
3~mJy beam$^{-1}$ for the NH$_3$(1,1) and NH$_3$(2,2) data respectively.

\subsection{Data reduction and fitting procedures}
\label{datared}

The reduction of the IRAM-30m telescope and PdBI data has been carried
out with the GAG-software developed at IRAM and the Observatoire de Grenoble.
In the next subsections we briefly outline the fitting procedures adopted 
to analyze the \ace\ and \AMM\ spectra.

\subsubsection{\ace\ fitting procedure}

\ace\ is a symmetric-top molecule with small dipole moment ($\mu=0.75$ Debye). 
Its rotational levels are described by two quantum numbers: $J$,
associated with the total angular momentum, and $K$, its projection on
the symmetry axis. Such a structure entails that for each $J+1\rightarrow J$
radiative transition, $J+1$ lines with $K\leq J$ can be seen (for a detailed
description see Townes \& Schawlow \cite{townes}). In our observations the
bandwidth covers all components for the (6--5) and (8--7) transitions, and 
up to the $K=10$ component for the (13--12) line. However, we detected only 
lines up to $K=3$.

In order to compute
the line parameters, we have performed Gaussian fits to the observed spectra
assuming that all the $K$ components of each $J+1\rightarrow J$ transition
arise from the same gas. Hence, they have the same LSR velocity
and line width. Then, in the fit procedure we fixed the line separation
in each spectrum to the laboratory value and we assumed the line widths
to be identical. In Sect.~\ref{phipar}
we will use the line intensities derived from this procedure to estimate the
kinetic temperature of the emitting gas. 

\subsubsection{\AMM\ fitting procedure}

The \AMM\ (1,1) and (2,2) inversion lines show hyperfine structure
(see i.e. Townes \& Schawlow \cite{townes}). To take into account this
structure, we fitted the lines using METHOD \AMM (1,1) and METHOD \AMM (2,2) 
of the CLASS program.
In this case the fit to the \AMM\ lines is performed assuming
that all components 
have equal excitation temperatures, that the line separation is fixed at the
laboratory value and that the linewidths are identical. This method
also gives an estimate of the total
optical depth of the lines using the intensity ratio between the 
different hyperfine components.

\section{Results}
\label{res}

\subsection{IRAM-30m observations}
\subsubsection{\CII\ and \CIII\ spectra}
\label{spectra}

In Fig. 1 we show the \CII\ and \CIII\ spectra taken at the central position in
the maps, corresponding to the peak of the sub-mm emission detected by 
Molinari et al. (\cite{mol00}). Single-pointing spectra of the \ace\ (6--5),
(8--7) and (13--12) lines are shown in Fig. 2. The \CII\ (1--0) and 
\CIII\ (1--0) spectra towards the central position clearly indicate the 
presence of two velocity components, centered at $\sim-50.5$ \kms and at
$\sim-47.8$ \kms , respectively (see Fig.~1).
The component with higher velocity appears very strong especially in the 
\CII\ (1--0) spectrum. We have fitted the two components with Gaussians.
In particular, the \CIII\ (1--0) line has hyperfine structure: thus, we have
performed Gaussian fits taking into account this structure by using
METHOD HFS of the CLASS program. This procedure also gives an estimate
of the line optical depth: we will discuss this point in Sect.~\ref{optic}.
The line parameters of the two components at the peak position are given in
 Table~\ref{linepar}.
Peak velocity ($V_{\rm lsr}$), full width half maximum ($\Delta v$) and 
integrated intensity of the line ($\int T_{\rm MB}{\rm d}v$) are given in
Cols. 3, 4 and 5, respectively. In Col. 2 we 
also give the 1$\sigma$ rms of the spectra. In the \ace\ observations only 
the lower velocity component is seen.
For these lines, we have performed Gaussian fits as explained in 
Sect.~\ref{datared}. The results are given in Table~\ref{aceres}.
 
\subsubsection{\CII\ and \CIII\ integrated intensity maps}
\label{irammap}
\begin{figure}
\centerline{\includegraphics[angle=0,width=7.5cm]{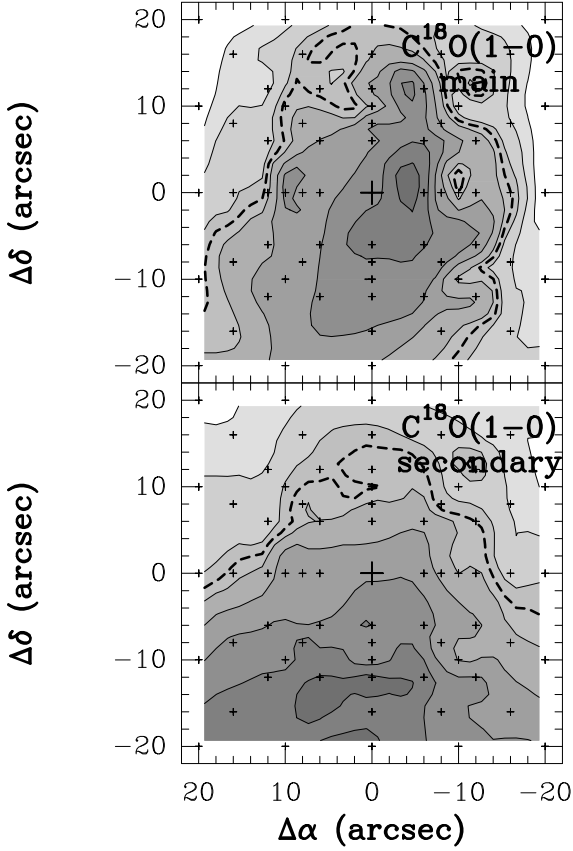}}
\caption{Top panel: IRAM-30m map of the main component of the \CII\ (1--0) 
line obtained integrating
between $-52.5$ and $-49.0$ \kms . Contour levels go from 
$\sim1.5$ K km s$^{-1}$ 
(corresponding to $\sim 3\sigma$), to the maximum value of 
$\sim$4 K km s$^{-1}$
 in steps of 0.5 K km s$^{-1}$ ($\sim1\sigma$). The thick, dashed contour indicates the half maximum 
power contour. Bottom panel: same as Top for the secondary component of the 
\CII\ (1--0) line, integrated between $-49.0$ and $-45.0$ \kms . Contour 
levels range from 1 K \kms\ ($\sim 3\sigma$) to 3.3 K \kms\ in steps of 0.3
K \kms\ ($\sim 1 \sigma$).}
\label{c1830m}
\end{figure}

\begin{figure}
\centerline{\includegraphics[angle=-90,width=7cm]{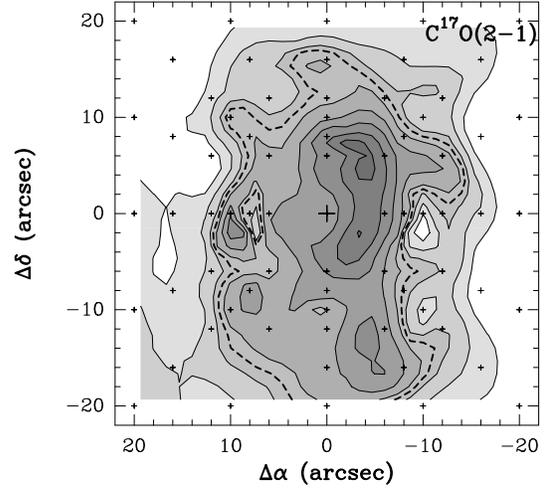}}
\caption{Same as Fig.~\ref{c1830m} for the main component of the \CIII\ 
(2--1) line, integrated between $-52.5$ and $-49.0$ \kms . Contour levels
 go from $\sim1$K km s$^{-1}$ ($\sim 3\sigma$ level), to 3.3 K km
 s$^{-1}$ in steps of $1\sigma$.}
\label{c17230m}
\end{figure}

Maps of the spectral component at $-47.8$ \kms\ show that the \CII\ (1--0)
(Fig.~\ref{c1830m}) and the \CIII\ (2--1) lines peak at (0\asec ,$-15$\asec ) 
from the central position of \II , suggesting that this emission is unrelated 
to this source. The presence of a 
secondary source to the South was already found by 
Molinari et al. (\cite{mol98b}) in the HCO$^{+}$(1--0) line.
Only the component centered at $-50.5$ \kms\ represents the
emission arising from \II : hereafter we shall refer to this as the ``main'' 
component. The secondary 
component, centered at $-47.8$ \kms , will not be discussed further.
\begin{table*}
\begin{center}
\caption[] {Line parameters of CO isotopomers at the peak position (IRAM-30m
observations). 
All line parameters have been obtained from
a two-Gaussian fit, except for the \CIII\ (1--0) line, for which we have used
METHOD HFS of the CLASS program to take into account the hyperfine structure 
(see text).}
\label{linepar}
\begin{tabular}{ccccc}
\hline \hline
Line  &  rms  &  $V_{\rm lsr}$ & $\Delta v_{1/2}$ & $\int T_{\rm MB}{\rm d}v$  \\
     & (K)  &  (\kms\ )  &   (\kms\ )   & (K \kms\ )   \\
\hline
\CII\ (1--0) &  0.065  &  $-50.0$  & 2.38  & 3.73 \\
             &         &  $-47.8$  & 1.20  & 2.59 \\
\CIII\ (1--0) &  0.064  &  $-49.9$  &  2.08 &  1.18  \\
              &         &  $-47.5$  &  1.12 &  0.41  \\ 
\CIII\ (2--1) & 0.18  & $-49.8$ & 2.93  & 5.06 \\
              &       & $-47.3$ & 0.90  & 0.69 \\
\hline
\end{tabular}
\end{center}
\end{table*}
\begin{table*}
\begin{center}
\caption[] {\ace\ line parameters (IRAM-30m observations)}
\label{aceres}
\begin{tabular}{lccccccc}
\hline \hline
Line  & rms & $v_{\rm LSR}$ &  $\Delta v_{1/2}$ &  \multicolumn{4}{c}{$\int T_{\rm MB}{\rm d}v$ (K \kms)}  \\
\cline{5-8}
          & (K) &  (\kms)  &  (\kms)     & $K=0$ & 1 & 2 & 3  \\
\hline
\ace\ (6--5) & 0.011  & $-50.3$ & 2.48   & 0.37 & 0.29 & 0.13 & 0.07 \\
\ace\ (8--7) &  0.024 & $-50.5$  & 2.64   & 0.41 & 0.40 & 0.22 & 0.13 \\
\ace (13--12) & 0.032 &  $-50.5$ & 2.57  &  0.75 & 0.71 & 0.41 & 0.30 \\
\hline
\end{tabular}
\end{center}
\end{table*}
The integrated intensity maps of the main component are shown in 
Figs.~\ref{c1830m} and~\ref{c17230m}. 
We do not show the \CIII\ (1--0) integrated map because the line is faint
over the whole map (S/N$\leq 3\sigma$ in many points).
The first clear result is that the emission looks quite flat and without 
an obvious peak towards \II . This is especially evident in the 
\CIII\ map.  
Also, in the \CII\ (1--0) and \CIII\ (2--1) maps, the half maximum power 
contour is elongated in the N-S direction: this could
be due to the presence of the secondary component in the spectra. In fact, it
is difficult to separate the two components, especially
 in the southern part of the map
where the overlap of the two lines is more pronounced. In an attempt
to derive the size of the emtting region in each line, we 
have estimated the angular diameter considering only the northern
part of the maps (that with positive offset in $\delta$), which is less 
affected by the secondary component:
we find an average diameter of the emitting region $\sim22$\asec\ for
the \CII\ (1--0) line, and $\sim 18$\asec\ for the \CIII\ (2--1) line. After 
correcting for the beam size (see Table~\ref{30mobs}), one finds a 
deconvolved angular diameter of $\sim18$\asec\ (corresponding to $\sim 0.45$
 pc) for the \CII\ (1--0) line and of $\sim15$\asec\ ($\sim 0.30$ pc) for 
the \CIII\ (2--1) line. 

In conclusion, from the 30-m maps of \II\ in the \CII\ and \CIII\ lines,
one cannot see an intensity peak towards the nominal 
position of the core, which instead is clearly seen 
by Molinari et al. (\cite{mol98b}) in different molecular transitions, as 
well as in our PdBI maps in the same lines, as shown
in Sect.~\ref{pdbobs}. We will further discuss this point in Sect.~\ref{depletion}.

\subsubsection{Optical depth of the lines}
\label{optic}

The availability of two rotational transitions of
the same molecular species (namely \CIII\ (1--0) and (2--1)) allows us to 
derive a map of the kinetic temperature, as we will explain in 
Sect.~\ref{tempco}.
However, this requires knowledge of the optical depth of such lines.
Since we have also observed the \CII\ (1--0) line, it is possible to
estimate the opacity of the \CIII\ (1--0) line from the ratio of two spectra,
assuming equal temperatures and beam
filling factors, and an abundance ratio of 3.7 for $^{18}$O/$^{17}$O (see 
Penzias \cite{penzias}, Wilson \& Rood \cite{wilson}).
We find an intensity ratio typically between 3
 and 4, with a mean value of $\sim 3.5$, which implies that the
\CIII\ (1--0) line is optically thin.

For \CIII , it is possible to derive the optical depth from the measure of 
the relative intensities of the hyperfine components.
 Unfortunately, in our spectra we are not able to resolve the 
hyperfine structure, hence the optical depth derived in this way is not
reliable. Concerning the 
\CIII\ (2--1) line, we cannot measure the optical
depth directly from our observations. However, in LTE conditions, one
can demonstrate that up to
temperatures as high as $\sim40$ K, the optical depth of the (2--1) line is
$\leq$ 3 times that of the (1--0) line. Since the maximum value
of the opacity of the (1--0) line derived from the line ratio is 
$\tau\sim 0.18$, the (2--1) line must have an 
optical depth less than $\sim 0.5$. Thus, when in Sect.~\ref{tempco}
we will derive the kinetic temperature from the ratio between the \CIII\ 
(2--1) and (1--0) lines, we
will assume that the lines are optically thin. 


\subsection{JCMT observations}

\begin{figure}
\centerline{\includegraphics[angle=-90,width=7cm]{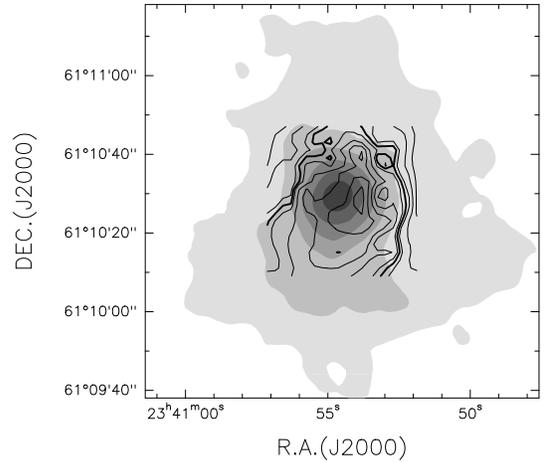}}
\caption{SCUBA map of \I\ at 850 $\mu$m (grey scale). Levels range from
0.3 (3$\sigma$) to 1.9 by 0.2 Jy beam$^{-1}$. The solid contours represent 
the \CII\ (1--0) map (main component) obtained with the IRAM-30m telescope 
(Fig.~\ref{c1830m}), with the 
thick line corresponding to the contour at half maximum.} 
\label{scuba_co}
\end{figure}

\II\ has been mapped in the sub-mm continuum at 850$\mu$m with SCUBA, with an
angular resolution of 15 \asec\ (Fig.~\ref{scuba_co}). The intensity
profile presents an unresolved compact central peak, and an extended halo
which surrounds the peak up to $\sim 100$ \asec . The figure also 
shows the integrated emission of the main component of the \CII\ (1--0)
line (upper panel of Fig.~\ref{c1830m}).
We can see that there is a reasonable match between the 850 $\mu$m 
emitting region and that traced by the \CII\ (1--0) line, although the 
corresponding peaks are not coincident.

\subsection{Plateau de Bure observations}
\label{pdbobs}

We present here the results of molecular line and millimeter continuum
observations carried out with the Plateau de Bure Interferometer.

\subsubsection{Molecular lines}
\label{lines}

\begin{figure}[h!]
\centerline{\includegraphics[angle=0,width=7.5cm]{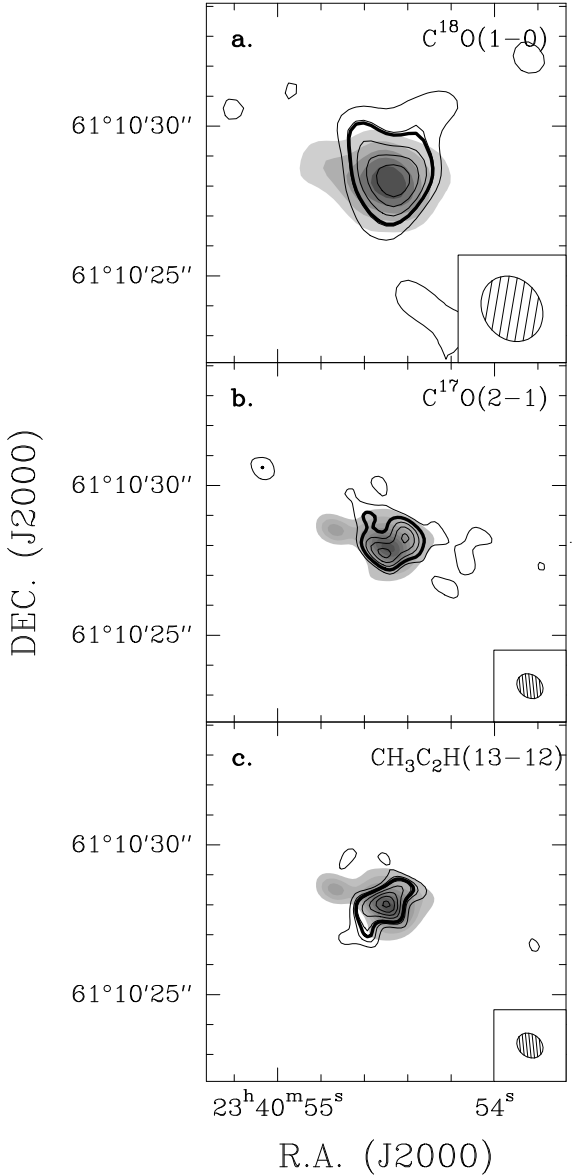}}
\caption{Maps obtained with the PdBI.
{\bf a}: C$^{18}$O(1--0) map integrated over the velocity range
($-49$,$-52$) \kms . Contour levels range from 0.05 ($\sim3\sigma$) to 0.14 by
0.02 Jy beam$^{-1}$. The thick contour indicates the half maximum power 
contour. The spectral rms is $\sim 0.014$ Jy beam$^{-1}$. The grey-scale image shows the 3~mm continuum emission, ranging
from $\sim1.9$ ($\sim4 \sigma$) to $7.4$ by 0.8 mJy beam$^{-1}$. The 
ellipse at the bottom right represents the synthesised beam. 
{\bf b}: same as {\bf a} for the \CIII\ (2--1) line. The map has been integrated over
 ($-47$,$-52$) \kms . Contour levels range from 0.03 to 0.09 by 0.015 Jy beam$^{-1}$.
The grey-scale image represents the 1.3~mm continuum emission, whose
countour levels range from 6 to 24 by 2.5 mJy beam$^{-1}$. The spectral rms is $\sim
0.012$ Jy beam$^{-1}$.
{\bf c}: same as {\bf b} for the \ace\ (13--12) line, integrated under the K=0
and 1 lines. Contour levels range from 0.03 to 0.1 by 0.01 Jy beam$^{-1}$. The 
spectral rms is $\sim0.013$ Jy beam$^{-1}$}
\label{mappepdb}
\end{figure}

All molecular transitions observed with the IRAM-30m telescope have been also
imaged with the PdBI, with the exception of the \ace\ (6--5) and (8--7) lines.
In Fig.~\ref{mappepdb} we show the maps obtained integrating the 
emission under the
\CII\ (1--0), \CIII\ (2--1) and \ace\ (13--12) lines. The secondary 
component is not detected, except perhaps in
\CII\ (1--0), in which the shape of the red wing of the line 
might be affected by the secondary component. This is likely due to the fact
that the secondary component is more extended, thus its emission has been
resolved out. 

We do not show the \CIII\ (1--0) map
because this is very noisy, and the signal is undetected up to a
$3\sigma$ level of $\sim 36$ mJy beam$^{-1}$.
The \CII\ (1--0), \CIII\ (2--1) and \ace\ (13--12) emission is detected with
a good signal-to-noise ratio. We can clearly see the presence of a 
central compact core coincident with \II . In the \CIII\ (2--1) and \ace\ 
(13--12) lines, the source 
shows a complex morphology, while it looks more simple in the 
\CII\ (1--0) line. This is consistent with the different angular
resolution: at 1.3~mm the beam size is $\sim2$ times smaller, allowing us to
better resolve the core structure. 
In particular, it can be noted that the \CIII\ (2--1) emission is 
marginally depressed at the peak position of the continuum and the \ace\
(13--12) line. This might be suggestive of CO depletion in the inner region
of the core (see discussion in Sect.~5.2).

We have estimated 
the diameters of the emitting regions from the half maximum power contours 
shown in Fig.~\ref{mappepdb}:
if we assume the source to be Gaussian, we can 
derive the angular diameter after beam deconvolution, as
given in Col. 2 of Table~\ref{tmass}.

\begin{figure}
\centerline{\includegraphics[angle=0,width=6cm]{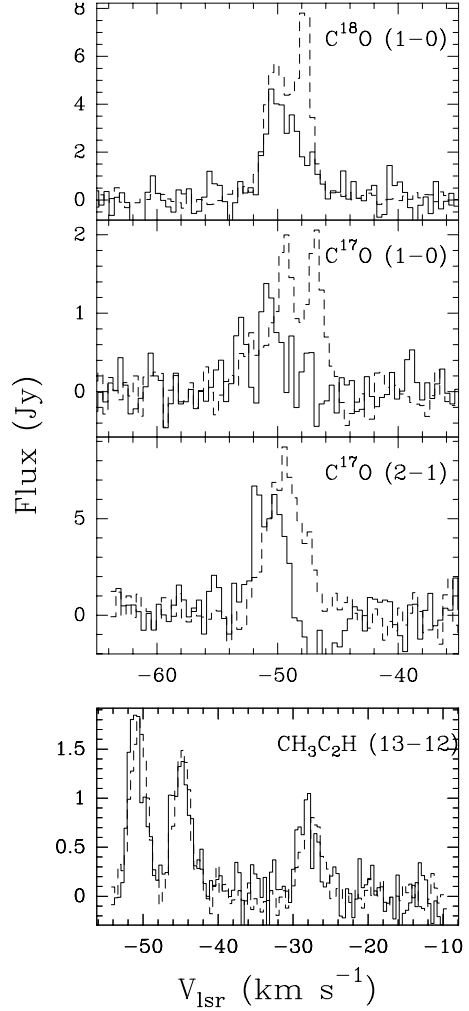}}
\caption{Flux density comparison between IRAM single-dish (dashed line) spectra
and Plateau-de-Bure interferometric spectra (solid line). The interferometric
flux densities have been multiplied by a factor of 10 for the CO isotopomers,
and by a factor of 3 for the \ace\ lines. The velocity of the \ace\
spectrum is computed with respect to the frequency of the line with $K=0$.}
\label{pdb-30m}
\end{figure}

In Fig.~\ref{pdb-30m} spectra of the \CII\ (1--0), \CIII\ (1--0) and (2--1) 
and \ace\ (13--12) lines are shown, that have been obtained integrating 
the emission
arising within the $4\sigma$ level inferred from the maps of 
Fig.~\ref{mappepdb}.
The availability of IRAM-30m spectra allows a comparison between 
interferometric and single-dish data. This comparison is also shown in 
Fig.~\ref{pdb-30m}, where the single-dish spectra have been resampled with 
the same resolution in velocity as the PdBI spectra, namely $\sim0.4$ \kms .
The superposition of the spectra clearly demonstrates that a large
fraction of the extended emission is filtered out by the interferometer.
 This is especially evident in the \CII\ (1--0) line, for which the flux 
measured
 with the PdBI is $\sim11$ times less than those obtained with the IRAM-30m 
telescope. This is also true
for the \CIII\ (2--1) line, for which the intensity ratio is $\sim10$.
In the \ace\ (13--12) transition this effect is smaller as the single-dish 
flux is $\sim3$ times larger. As noted earlier, the secondary 
component is resolved out in the PdBI spectra.

Looking at the \CIII\ (2--1) spectrum 
of Fig.~\ref{pdb-30m}, the line profile seen by the interferometer seems
to peak at a slightly blue-shifted velocity with respect to the 30-m
spectrum. We believe that this is due to the fact 
that the S/N ratio is poor in the PdBI spectrum, hence the peak velocity,
and the other line parameters, are affected by large uncertainties. 
In fact, the offest observed ($\sim 0.5$ \kms ) lies within the
$3\sigma$ uncertainty on the line peak velocity ($\sim0.45$ \kms )
obtained from a Gaussian fit.

\subsubsection{Millimeter continuum}
\label{micont}
\begin{figure}
\centerline{\includegraphics[angle=0,width=6cm]{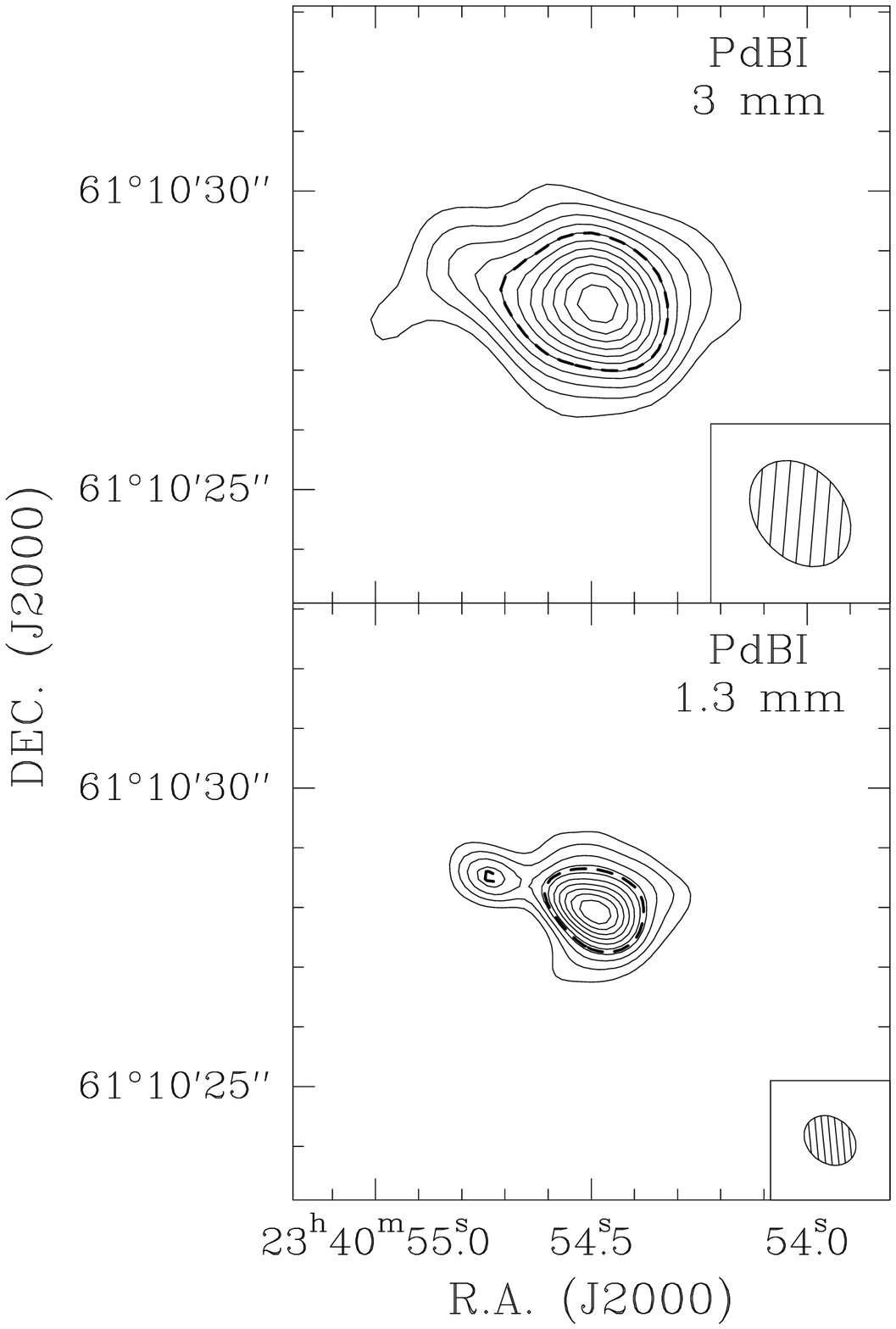}}
\caption{Top panel: map of the 3~mm continuum emission. The rms level is $\sim 0.6$ mJy beam$^{-1}$ and the signal-to-noise ratio is $\sim15$. Contour levels range
from 1.6 to 7.8 by 0.6 mJy beam$^{-1}$. The thick, dashed contour represents the half
maximum power contour. Bottom panel:same as top for 1~mm continuum
emission. The rms is $\sim 2$ mJy beam$^{-1}$ at 1.3~mm and the 
signal-to-noise ratio is $\sim$14. Contour levels range from 6 to 24 by 2 mJy 
beam$^{-1}$. The ellipses in
the bottom right are the synthesized beams.}
\label{contpdb}
\end{figure}

With the PdBI we detect a compact central core, as can be seen from 
Fig.~\ref{contpdb}, where we show the 3~mm and 1.3~mm 
continuum maps. The 3~mm map shows a compact core with an
elongated shape along the E-W direction. This shape is better resolved at 
1.3~mm, revealing a
secondary faint peak of emission offset by $\sim+2$ \asec\ in R.A. from the 
main peak. This secondary peak is not detected in any of the molecular tracers.
The angular diameters are 1.5\asec\ and 1.1\asec\ at 3~mm and 1.3~mm,
respectively.
Measured flux densities $F_{\nu}$ are 0.0124 and 
0.148 Jy at 3 and 1.3~mm, respectively, obtained by 
integrating over the solid angle down to the 3$\sigma$ level. 
The flux density of the secondary source at 1.3~mm is $\sim5\%$ of 
the total, namely 0.008 Jy.

We can compare the 
continuum flux density measured with the PdBI and with SCUBA, to estimate the 
amount of the flux density filtered out by the interferometer. Using a 
spectral 
index for the dust emissivity of 4 (see Sect.~\ref{contspec}), we have 
extrapolated the flux density observed at 850 $\mu$m with 
SCUBA, to 1.3~mm. This has been
smoothed to an angular resolution of $\sim 22$\asec , corresponding 
to the primary beam of the PdBI at 1.3~mm. The flux density measured in this
map towards the core \II\ is $\sim 0.28$ Jy. We estimate in $\sim 0.11$ Jy 
the flux density to arise from the halo: this is filtered out by the 
interferometer, and thus the flux density that we should observe with the PdBI
is $\sim 0.17$ Jy. Taking into account the uncertainties 
that affect our measurements
(in particular the combined calibration errors of both SCUBA and PdBI, 
which are at least of the order of 30$\%$), the 
0.15 Jy observed at 1.3~mm with PdBI is consistent with 
what is expected. 
Thus, within the errors, it is plausible that with the PdBI we observe all 
the continuum flux arising from the core \II .

\subsection{Very Large Array \AMM\ maps}
\label{amm}

\begin{figure*}[h*]
\centerline{\includegraphics[angle=-90,width=13cm]{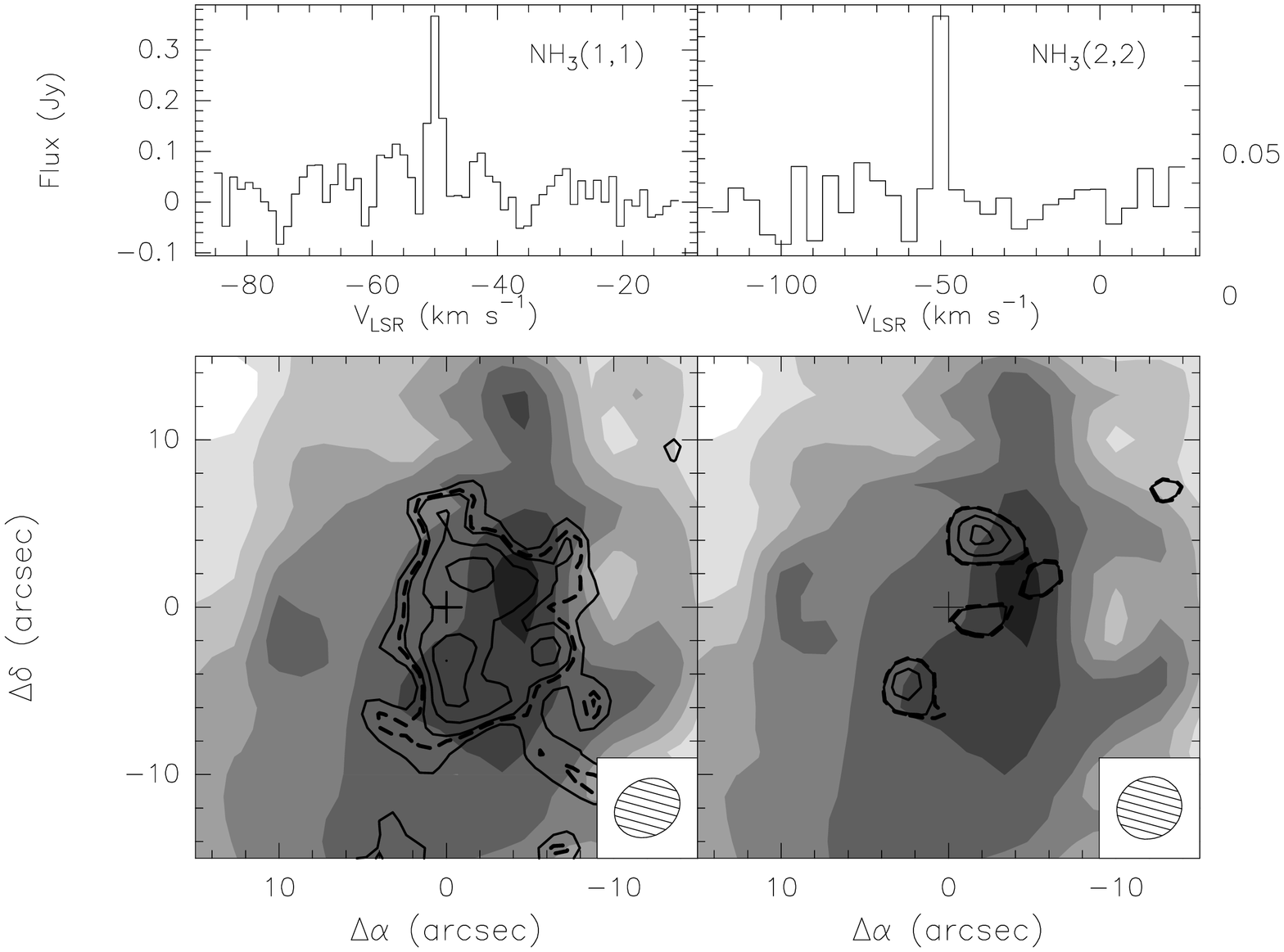}}
\caption{Left panel: \AMM  (1,1) integrated map over the velocity range ($-48$,
$-52$) \kms\ . Contour levels range from 0.012 ($\sim3\sigma$) to 0.024 by 
0.004 Jy beam$^{-1}$. The channel spacing of the integrated spectrum (shown
over the map) is 
$\sim1.13$ \kms . The grey-scale represents the \CII\ (1--0) main line emission
(Fig.~\ref{c1830m}). The cross indicates the core position detected with
the PdBI. Right panel: same as Left for \AMM (2,2). Contour levels 
range from 0.012 to 0.025 by 0.004 Jy beam$^{-1}$. The channel spacing is $\sim4.9$
\kms . The ellipses in 
the bottom-right represent the VLA synthesised beam.}
\label{nh3fig}
\end{figure*}

\begin{figure}[h]
\centerline{\includegraphics[angle=0,width=7cm]{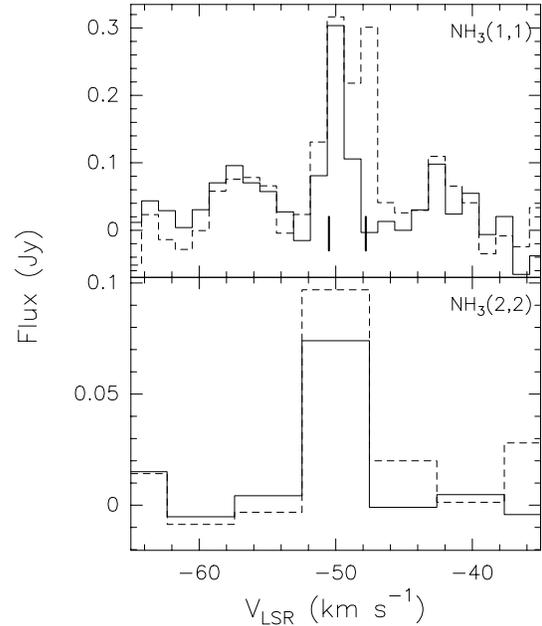}}
\caption{Flux density comparison between Effelsberg-100m (dashed line) and VLA 
(solid line) spectra of the \AMM (1,1) and (2,2) lines. VLA fluxes have
been multiplied by a factor of 2. The thick vertical lines under the
\AMM (1,1) spectrum indicate the position of the two components. The 
Effelsberg spectra have been resampled to the channel spacing of the VLA 
spectra.}
\label{eff_vla}
\end{figure}

With the VLA, we have mapped the \AMM\ (1,1) and (2,2) 
inversion transitions. The corresponding maps and integrated spectra are
 shown in Fig.~\ref{nh3fig}. In both maps, the \AMM\ emission is
superimposed on the \CII\ (1--0) map. As for the CO isotopomers 
observed with the IRAM-30m telescope, the emission does not peak towards
the position of the core detected at 3~mm and 1.3~mm with the PdBI. 
Although the signal-to-noise ratio is
poor in both maps (see Fig.~\ref{nh3fig}), one can see that the 
emission arises from a ``clumpy'' structure, and it peaks at $\sim5$\asec\
 ($\sim 0.12$ pc) from the core position. 
In Sect.~\ref{depletion} we will discuss
this result by comparing the \AMM\ maps with those obtained in the other
tracers. Comparison between Effelsberg-100m (Molinari et al. \cite{mol96})
and VLA observations are shown
in Fig.~\ref{eff_vla}. For the main component (centered at $-50.5$ \kms ),
with the VLA we lose about the 50$\%$ of the total flux density seen with the 
single-dish telescope; for the \AMM\ (1,1) line, this occurs for all the 
hyperfine lines. Finally, we completely resolve out the secondary 
component (centered at $-47.8$ \kms) as in the other molecular tracers observed
with the PdBI .

 
\section{Derivation of physical parameters}
\label{phipar}

Our goal is to investigate the hypothesis of Molinari
et al. (\cite{mol98b}), namely that \II\ is a high-mass protostellar object,
by examining
the physical parameters of the core. First, we will analyze the continuum
spectral energy distribution (SED). Then, from the 
PdBI map of the \ace\ (13--12) line, we will compute the core temperature; 
finally, from continuum and line emission we will obtain the mass 
and the H$_{2}$ volume density of the core. 

\subsection{Continuum SED and source luminosity}
\label{contspec}
\begin{figure}
\centerline{\includegraphics[angle=-90,width=8cm]{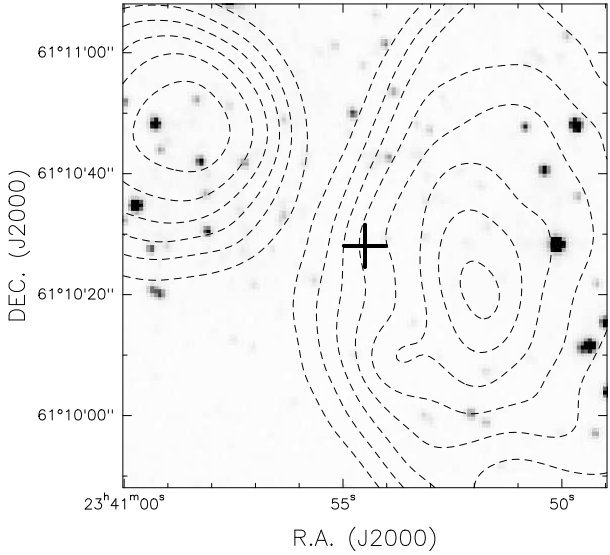}}
\caption{Image: \I\ observed at 2 $\mu$m with the IRCAM of the Palomar 
telescope in the Ks band (Molinari priv. comm.), which shows the cluster
of stars around the core \II\ (indicated by the cross at map center). The 
dashed contours represent the emission of the two extended radio sources 
detected at 3.6~cm with the VLA: contour levels range between 0.3 and 3 by 
0.3 mJy beam$^{-1}$ (Molinari et al. \cite{mol02}).}
\label{ks}
\end{figure}
\begin{figure}
\centerline{\includegraphics[angle=-90,width=8cm]{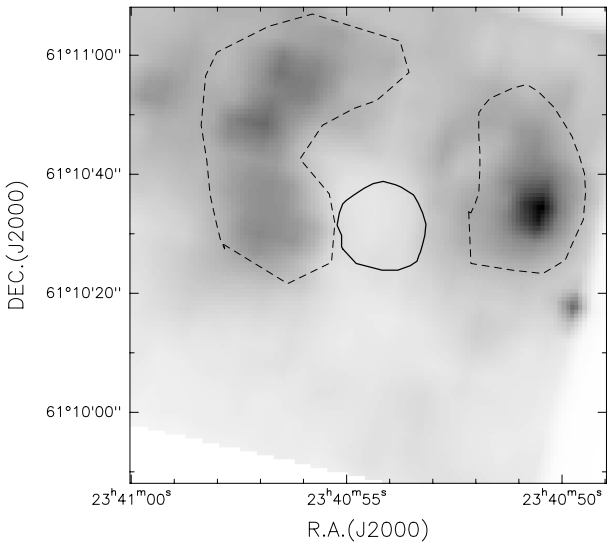}}
\caption{Image of \I\ at 15~$\mu$m obtained with ISOCAM (Molinari et 
al.~\cite{mol98b}). The thick polygon represents the 3$\sigma$ level of the 
\HCOp\ (1--0) line, and it identifies the core region. The dashed polygons 
indicate the two clusters of stars which lie around \II , called ``ring'' in
the text.}
\label{poligoni}
\end{figure}

\begin{figure}
\centerline{\includegraphics[angle=0,width=8cm]{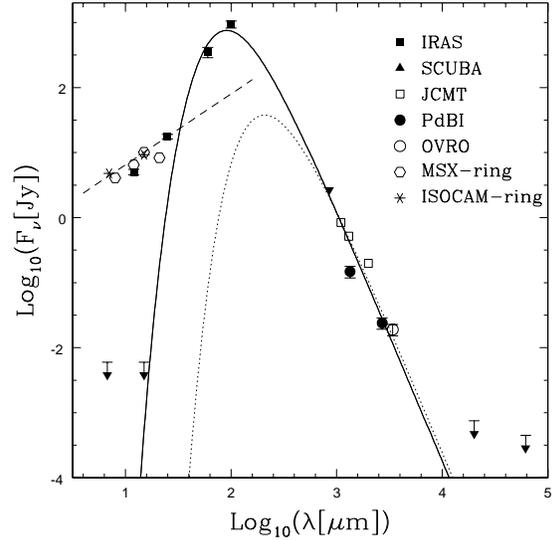}}
\caption{Continuum SED of \II . Open 
hexagons and stars indicate MSX and ISOCAM
flux densities respectively measured integrating over the ``ring'' region 
of Fig.~\ref{poligoni}. The filled triangle represents the SCUBA measurement. 
Filled squares, open squares, filled circles and open
circles indicate respectively IRAS, JCMT, PdBI and OVRO data (see also
Fig. 2 of Molinari et al. \cite{mol98b}). The arrows on the bottom left
and bottom right respectively indicate the ISOCAM and VLA upper limits 
($3\sigma$ level) of the core flux density.
The solid line represents a grey-body fit, 
with dust temperature $T=40$ K and dust absorption coefficient 
$\beta\propto\nu^{1.9}$, while the dotted line is a grey-body fit with the
same $\beta$ and $T=15$ K.
The dashed line corresponds to a power-law extrapolation
of the flux density arising from the ``ring'' region.}  
\label{sed}
\end{figure}

In discussing the continuum SED, it is important to distinguish between
two ``sub-regions'': 
\begin{itemize}
\item the core \II ; 
\item a stellar and diffuse emission from a cluster surrounding the core
(see Fig.~\ref{ks}).
\end{itemize} 

The cluster is detected as a patchy ``ring'' 
at FIR wavelengths with ISOCAM (Fig.~\ref{poligoni}) and it seems to
overlap with two extended sources detected at 3.6~cm with the VLA. From the
radio flux Molinari et
al. (\cite{mol02}) have estimated the spectral type of the ionizing stars
to be B0.5.

In Fig.~\ref{sed} the continuum SED of \II\ is shown. With respect to
Fig. 2 of Molinari et al. (\cite{mol98b}), Fig.~\ref{sed} contains more
data, namely PdBI, IRAS, JCMT, OVRO, ISOCAM and MSX \footnote{ MSX images have been taken from the
on-line MSX database http://www.ipac.caltech.edu/ipac/msx/msx.html}
. Since no 7 and 15
$\mu$m emission is detected with ISOCAM towards the source, we take the 
$3\sigma$ level as an
upper limit at these wavelengths. The same was done for the VLA measurements
at centimeter wavelengths.

We performed a grey-body fit to the continuum SED of Fig.~\ref{sed}
(solid line).
The resulting bolometric luminosity is the same as in Molinari et al.
(\cite{mol98b}), namely $L\sim1.6\;10^{4}\; L_{\odot}$. The slope of
the millimetric part of the spectrum is fitted assuming a dust
opacity $k_{\nu}=k_{230 {\rm GHz}}(\nu/230(\rm GHz))^{\beta}$, 
where $k_{230 {\rm GHz}}=0.005{\rm\; cm^{2} g^{-1}}$, which
assumes a gas-to-dust mass ratio of 100 (see
Preibisch et al.~\cite{preib}). 
The best fit is obtained for $F_{\nu}\propto\nu^{3.9}$. Therefore, the 
best fit opacity index is $\beta\simeq 1.9$. The other best fit parameters
are: source temperature of 40 K, diameter of 8\asec and a mass of $\sim 400M_{\odot}$. Since most of the luminosity is due to the IRAS 60 and 100 $\mu$m
flux densities, it is
important to assertain the origin of the IRAS flux densities. In fact,
the IRAS beam is much larger than the source diameter ($\sim2$ arcmin at
100 $\mu$m), thus the IRAS flux densities may arise both from \II\ and 
from the
cluster of stars surrounding it (the ``ring'' in the 15 $\mu$m ISOCAM map of
Fig.~\ref{poligoni}). Hence, the source
luminosity estimated before ($L\sim1.6\;10^{4}\;{\rm L_{\odot}}$) is to be 
regarded as an upper limit for the luminosity of \II . However, we know
from the JCMT and interferometric maps
that the flux densities at 850 $\mu$m and at longer wavelenghts certainly 
come from 
\II . Thus, we performed a grey-body fit using only these data (dotted line
of Fig.~\ref{sed}): the corresponding grey-body fit gives a luminosity of
$\sim 150\;L_{\odot}$ which represents a lower limit.
The two limits thus obtained are very different, but we believe that the source
luminosity is closer to the upper limit.
In fact, the MSX image at 21 $\mu$m shows that the emission at this
wavelength arises from the ``ring'', hence very likely this region is
also responsible for the 25 $\mu$m emission. However, as already 
discussed by Molinari et 
al. (\cite{mol98b}), it is unlikely that the ``ring'' 
contributes significantly to the IRAS flux densities at 60 and 100 $\mu$m, 
because
even a power-law extrapolation of the spectrum of the extended region
(dashed line in Fig.~\ref{sed}) gives flux densities at 60 and 100 $\mu$m much
lower (a factor of $\sim 10$) than the values observed.
We conclude that most of the emission at 60 and 100 $\mu$m is due to
\II , and therefore a luminosity of $\sim 10^{4}\;L_{\odot}$ is not very
far from the real value. We will further discuss this point in 
Sect.~\ref{discu_core}, on the basis of our \ace\ observations.

It is also worth noting that the 1.3~mm flux density measured with the PdBI 
is $\sim3$ 
times less than the value measured with the JCMT at the same wavelength.
In fact, as explained in Sect.~\ref{micont}, with the interferometer we
observe approximately all the flux density arising from the compact core, but 
we miss that coming from the extended halo, which we do detect with the JCMT.
Finally, it is unlikely that the source associated with the
secondary peak detected in the 1.3~mm map contributes significantly to the
total observed luminosity, because its 
1.3~mm flux density is only 5$\%$ of that from the main core.

\subsection{Kinetic temperature and column density from \ace\ rotation 
diagrams}
\label{bolt}

From the \ace\ (13--12) spectra of Figs.~2 and~\ref{pdb-30m} we derive the kinetic 
temperature and the total column density of the molecule by means of the
Boltzmann diagrams method (see e.g. Fontani et al. \cite{fonta}). The 
fundamental assumption of the method is the local thermodynamic equilibrium
(LTE) condition of the gas. Such an assumption is believed to work very well
for \ace\ because of its low dipole moment (see Bergin et al. \cite{bergin}).
Under the further assumption of optically thin emission, the
column density $N_{\rm i}$ of the upper level $i$ of each transition is 
proportional to the integrated intensity of the line.
 
\begin{figure}
\centerline{\includegraphics[angle=0,width=7cm]{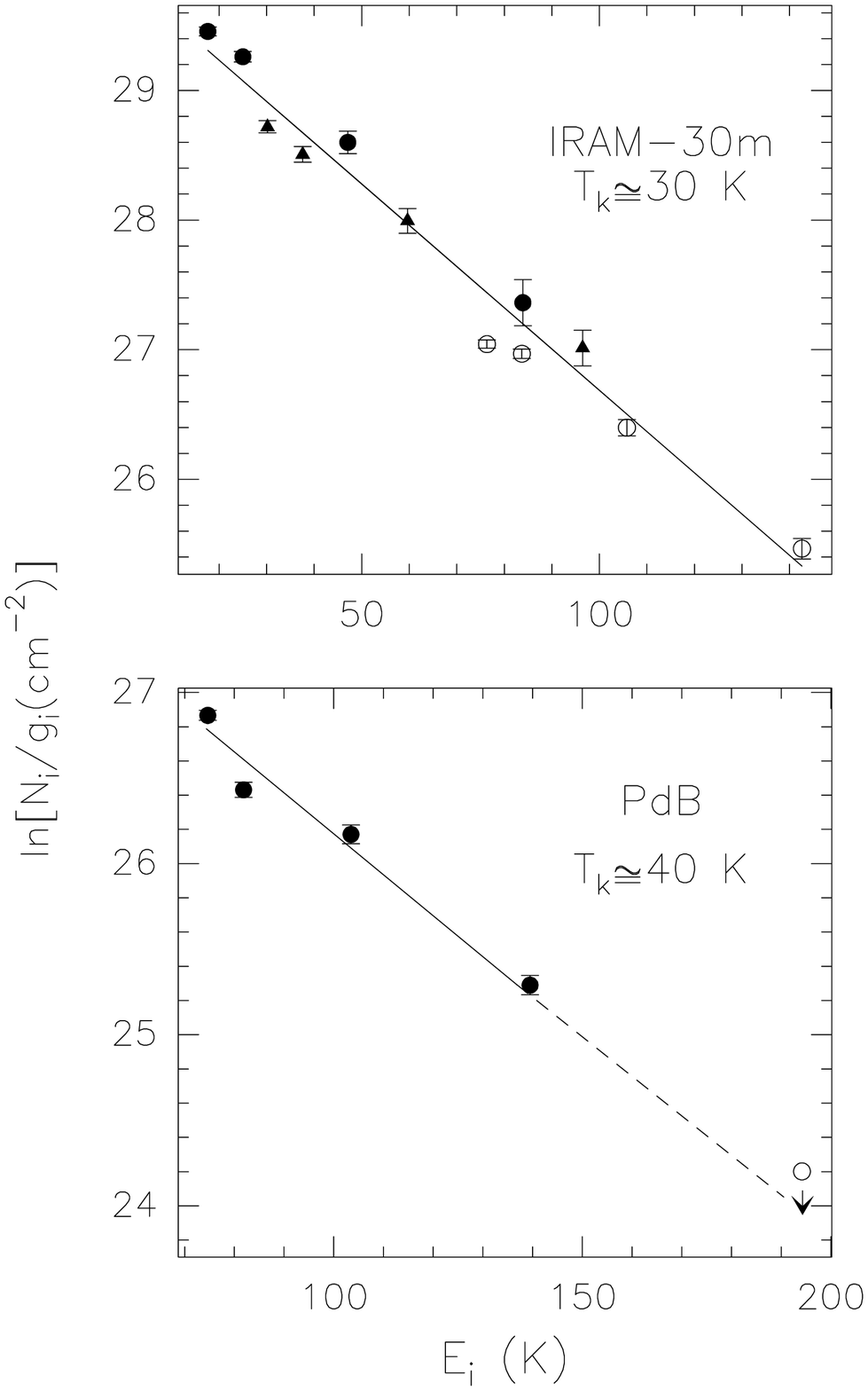}}
\caption{Top panel: rotation diagram inferred from \ace\ lines observed 
with IRAM-30m telescope. Filled circles, filled triangles and open circles
indicate respectively the (6--5), (8--7) and (13--12) transitions. The straight
line represents a least square fit to the data. Column densities are 
source-averaged values using a source diameter of $15$\asec\ (see Sect.
\ref{bolt}). Bottom panel: same as Top for \ace\ (13--12) observed with
PdBI. The assumed source size is $1.4$\asec . The open circle represents the
$3\sigma$ level of the spectrum, taken as an upper limit for the $K=4$ 
component.}
\label{bzplot}
\end{figure}

In Fig.~\ref{bzplot} Boltzmann diagrams are shown: the straight 
lines represent least square fits to the data, and the kinetic temperature and
the column density are related to the slope and the intercepta, respectively. 
The interferometric data
yield $T_{\rm k}\simeq42$ K and $N_{\rm tot}\simeq4\;10^{14}$ \cmq\ for the 
core. It is evident that such temperature is well below that of a HC.
In a recent paper, Thompson \& Macdonald (\cite{thompson}) have observed
rotational transitions of the CH$_{3}$OH molecule towards \I , from which
they derive that the kinetic temperature is likely to be smaller than
50 K, and therefore it is unlikely that \I\ might contain a HC. Our study 
strongly supports this idea and give a much more accurate
temperature estimate of the core.

With the IRAM-30m data we obtain {\bf $T_{\rm k}\simeq34$ K} and 
{\bf $N_{\rm tot}\simeq5\;10^{15}$} \cmq . In this case, the column densities
 $N_{\rm i}$ have been corrected for the beam filling 
factor. For this purpose, an estimate of the \ace\ emitting region is 
needed. Unfortunately, we have no direct estimate because no \ace\ single-dish
maps are available. However, we can use the source size estimated from
the SCUBA map, given in Sect.~4.5. 

The previous results are summarised
in Table~\ref{tk}. The errors on $T_{\rm k}$ and $N_{\rm tot}$ have 
been computed in two different ways: for the single-dish data we must 
take into
account the calibration error. This error affects simultaneously each line in 
the same bandwidth. Hence, in the diagrams, the effect of this
error is to shift the points in the same bandwidth (e.g. the (6--5) lines) 
by the same quantity. This error is $\sim 10\%$ for the (6--5) lines
and  $\sim 20\%$ for the (8--7) and (13--12). Hence, in order to compute the 
uncertainties on $T_{\rm k}$ and $N_{\rm tot}$, we have varied the values
of $N_{\rm i}$ by 10\% simultaneously for all the (6--5) K components, and
by 20\% for the (8--7) and (13--12). These variations have been made in
different directions.
The uncertainties in Table~\ref{tk} are the maximum difference between
the values thus obtained and the nominal ones.
For the PdBI estimates, we have data of a single band, and the
calibration error does not affect \Tk\ but only $N_{\rm tot}$.

Let us now discuss the two basic assumptions of the method, i.e. 
optically thin emission and LTE conditions of the gas.
The LTE assumption holds very well when the H$_{2}$ density is greater than
a critical value (see e.g. Spitzer \cite{spitzer}). For the observed
transitions the maximum critical density is $\sim 10^{5}$ \cmc , and
we shall see in Sect.~\ref{masses} that the inferred H$_{2}$ density is 
above this value, justifying the assumption of LTE conditions.
Concerning the optical depth of the lines, the excellent agreement between
the data and the linear fit of Fig.~\ref{bzplot} strongly supports the 
assumption of optically thin lines. In fact, large optical
depths are expected to affect mainly the transitions with lower excitation. 
Hence we should have a flattening of the Boltzmann diagram at lower energies, 
which is not seen in Fig.~\ref{bzplot}.

\subsection{Physical parameters from CO isotopomers}
\label{tempco}

In the previous section we have computed the kinetic temperature of the source
from the \ace\ lines. We can also estimate the temperature by means of 
the \CIII\
lines observed with the IRAM-30m telescope. Using a source size of $15$\asec ,
found in Sect.~\ref{irammap} from the \CIII\ (2--1) map, we have derived
the beam-averaged brightness temperatures for the \CIII\ (1--0) and (2--1)
lines. Then, from the integrated line intensities, and assuming optically thin
lines and LTE conditions, we find a ``Boltzmann relationship'' between the 
total column density of the upper level of each transition and the 
corresponding rotational energy, as
discussed above for the \ace\ lines. We have done this
for each point of the \CIII\ maps in which the \CIII\ (1--0) line
is stronger than the $3\sigma$ level. The resulting average kinetic 
temperature is $\sim 15$ K. 
It is important to stress that in deriving the kinetic temperature from 
\CIII\ line ratios we have assumed LTE conditions. Statistical
equilibrium calculations (see Wyrowski \cite{wyrowski}) show that this
assumption holds if $n_{\rm H_{2}}\geq10^{5}$ \cmc . Otherwise, the
temperature estimated in this way is to be regarded as a lower limit. Since 
we obtain
$n_{\rm H_{2}}\sim10^{5}$ \cmc\ over the whole map, we conclude that the
LTE assumption is satisfied in our case.

We obtain a total column density ranging from
$9\;10^{14}$ to $7\;10^{15}$ \cmq : the \CIII\ total column
density given in Table~\ref{tk}, namely $4\;10^{15}$ \cmq , is the mean 
value over the map.

\subsection{Physical parameters from \AMM\ (1,1) and (2,2)}
\label{tempamm}

The \AMM\ emission seen with the VLA arises from a region
whose diameter is ``intermediate'' between that of the core \II\ and that of
the \CIII\ and \CII\ total emitting region seen with the IRAM-30m telescope. 
The emission peak does not coincide with the continuum nor with
the \CII\ (1--0) line.
The \AMM\ (1,1) hyperfine structure allows us to estimate the optical depth
as explained in Sect. \ref{datared}: we obtain 
$\tau \ll 1$ for this line. For the (2,2), we have no direct estimates, but 
we can reasonably assume that for temperatures between $\sim10$ and 
$\sim40$ K, under LTE conditions, the (2,2) line is also optically thin.
The temperature 
and column density can be derived following the method by Ungerechts et al. 
(\cite{ung}). We obtain a temperature of $\sim 26$ K and a total column
density of 6 $10^{15}$ \cmq . We have 
assumed equal linewidth and beam filling factor for both transitions. The 
results are given in Table~\ref{tk}.

\subsection{Mass and density estimate}
\label{masses}

The grey-body fit to the spectrum of Fig.~\ref{sed} is obtained for a dust 
temperature of 40 K, a dust 
opacity $k_{\nu}\propto\nu^{1.9}$, an average source diameter of 8\asec ,  
($\sim0.2$ pc), and a mass of $\sim 400$ M$_{\odot}$. 

It is also useful to derive the mass of the core seen with the PdBI. From 
the 3~mm continuum flux density, using the core temperature
deduced from the \ace\ lines (namely $T=42$ K), $k_{\nu}\propto\nu^{1.9}$, 
and assuming optically thin dust emission,
we find that the mass contained inside $\sim$1\pas 5 is 
$\sim150$ M$_{\odot}$. Note that we have used the 3~mm flux density, 
rather than that at 1.3~mm, because we expect to miss less flux 
at this wavelength.

To compute the mass from the molecular line emission, we use the expression:
\begin{equation}
M_{\rm line}=\frac{\pi}{4} \frac{D({\rm cm})^{2} N_{\rm tot}({\rm cm^{-2}})
m_{\rm H_{2}}({\rm g})}{X}     \label{emcd}
\end{equation}
where $D$ is the source diameter, $m_{\rm H_2}$ the mass of the H$_2$
molecule, $X$ the abundance of the molecule relative to H$_2$,
and $N_{\rm tot}$ is the source averaged column density.
$X=1.2\;10^{-7}$, 
$X=3.4\; 10^{-8}$ and $X=2\;10^{-9}$ for the \CII , \CIII\ and \ace\ mean abundances,
respectively. The relative abundances of the CO isotopomers
have been computed from 
Wilson \& Rood (\cite{wilson}) for a galactocentric source
distance of 11 kpc. The \ace\ abundance is an average value from Fontani 
et al. (\cite{fonta}) and Wang et al. (\cite{wang}), who studied the \ace\
emission in massive star formation sites. The resulting values of 
$M_{\rm line}$ are reported in Col.~7 of Table~5.

From the line-widths of the observed transitions and the 
corresponding angular diameters in Table~\ref{tmass},
we can also derive the mass required for virial equilibrium: 
assuming the source to be spherical and homogeneous, neglecting contributions
from magnetic field and surface pressure, the virial mass is given by
(MacLaren et al. \cite{mclaren}):
\begin{equation}
M_{\rm VIR}(M_\odot)=0.509\,d({\rm kpc})\,\Theta_{\rm s}({\rm arcsec})\,\Delta v_{1/2}^{2}({\rm km/s})   \label{emvir}
\end{equation}
In Table~\ref{tmass} we show the resulting virial mass deduced from the \ace\ 
(13--12) lines observed with the PdBI and also from other lines. 

Finally, from $M_{\rm line}$ we derive the $\rm 
H_{2}$ volume density, $n_{\rm H_{2}}$. All masses and densities 
are given in Tables~\ref{tmass} and \ref{tmass2}. It is worth noting that the density 
of the central core is $\geq\;10^{7}$ \cmc : this value is well above 
the maximum critical density of the \ace\ lines observed, thus supporting
our LTE assumption and our derivation of the kinetic temperature. 
As usual, the main errors in estimating the masses are due to
the molecular abundances, the assumed dust opacity and the gas-to-dust
ratio, and it is very difficult to quantify the uncertainties for these
parameters.

One can see clearly that masses and densities derived from different tracers
differ very much from one another.
This is partly due to the fact that different tracers arise from different
regions. However, even for the same region, the mass derived from
the continuum emission is a few times that estimated from the line emission.
This is likely due to uncertainties in the assumed molecular abundances.
 We shall come back to this point in Sect.~\ref{depletion}. 

\begin{table*}
\begin{center}
\caption[] {Temperature and column density estimates. $\theta_{\rm s}$ and 
$D$ are the angular and linear source diameter, respectively.\Tk\ is the
kinetic temperature, and $N_{\rm tot}$ the total column density of the
corresponding molecular species}
\label{tk}
\begin{tabular}{c|c|c|c|c}
\hline
Tracer  &  $\theta_{\rm s}$ & $D$ & \Tk\  & $N_{\rm tot}$  \\
    & (\asec\ ) & (pc) & (K) &  (\cmq\ )  \\
\hline
\ace\ (PdBI) & 1.4  & 0.033 & 42$\pm$1 & $(1.8\pm0.2)\;10^{14}$ \\
\AMM\  (VLA) &  $\sim 5$ & 0.12 &  26$\pm$6  & $\sim6{\bf \pm2}\;10^{15}$ \\ 
\ace\ (30m)$^{*}$  &  18 & 0.43 &  34$\pm$6 & $(5\pm0.2)\;10^{15}$ \\ 
\CIII\ (30m)$^{*}$ &  15  & 0.35 & 15$\pm$5 & $\sim4{\bf \pm2}\;10^{15}$ \\
\hline  
\end{tabular}
\end{center}
$^{*}$ corrected for source size
\end{table*}

\begin{table*}
\begin{center}
\caption[] {Mass and $H_2$ column- and  volume- density estimates from 
lines. Molecular 
abundances relative to $H_2$ have been assumed equal to 1.2 $10^{-7}$, 3.4
$10^{-8}$ and $2\;10^{-9}$ for \CII , \CIII\ and \ace , respectively.
$N_{\rm H_2}$ and $\Delta v_{1/2}$ are the ${\rm H_2}$ column density and the FWHM of the lines,
respectively}
\label{tmass}
\begin{tabular}{cccccccc}
\hline \hline
 Tracer  & $\theta_{\rm s}$  & $D$ & $N_{\rm H_2}$& $\Delta v_{1/2}$  & \mvir\ & 
$M_{{\rm line}}$ & $n_{\rm H_{2}}$ \\
   & (\asec\ )   & (pc) & (\cmq\ ) & (\kms\ )  &   ($M_\odot$) &  ($M_\odot$) & (\cmc\ ) \\
\hline
\CIII\ (2--1) (PdBI) & 1.3 & 0.03 & 5.4 $10^{24}$ & 2.9 & 30 & 42 & 6.0 10$^7$ \\
\ace\ (13--12) (PdBI) & 1.4 & 0.033 & 3.3 $10^{24}$ & 3.0 & 32 & 15 & 1.7 10$^{7}$ \\
\CII\ (1--0) (PdBI) & 1.9 & 0.045 & 4.3 $10^{24}$ & 3.4 & 55 & 69 & 3.2 10$^7$ \\
HCO$^{+}$(1--0)($^{\dagger}$) & 7 & 0.16 & 1.66 $10^{24}$ & 3.3  & 200 & 370 & 3.4 $10^6$ \\ 
\CIII\ (2--1) (30m) & 15 & 0.35 & 1.1 $10^{23}$ & 2.9 & 314 & 110 & 1.0 10$^{5}$ \\
\CII\ (1--0) (30m) & 18 & 0.43 & 1.0 $10^{23}$ & 2.4 & 259 & 200 & 1.1 10$^{5}$ \\
\hline
\end{tabular}
\end{center}
($^{\dagger}$) From Molinari et al. (\cite{mol02})
\end{table*}

\begin{table*}
\begin{center}
\caption[] {Mass and $H_2$ column- and volume- density estimated from 
continuum observations}
\label{tmass2}
\begin{tabular}{cccccc}
\hline \hline
 Tracer      &   $\theta_{\rm s}$ & $D$ & $N_{\rm H_2}$ & $M_{\rm cont}$ & $n_{\rm H_{2}}$  \\
      &   (\asec\ ) & (pc) & (\cmq\ ) & ($M_\odot$) & (\cmc\ ) \\
\hline
 3~mm (PdBI)  &   1.5   & 0.036 & 1.5 $ 10^{25}$ & 150 & 14.0 $ 10^7$ \\
3.4~mm (OVRO$^{\dagger}$)    &   4 & 0.09 & 0.4 $ 10^{25}$ & 324 & 1.6 $10^7$ \\
SED ($^*$) & 8 & 0.20 & 1.5 $10^{24}$ & 400 &  2.5 $10^{6}$ \\
850 $\mu$m (SCUBA)   & 18 & 0.43 & 3.6 $10^{23}$ & 520 &  2.8 $10^5$ \\
\hline
\end{tabular}
\end{center}
$^{\dagger}$ From Molinari et al. (\cite{mol98b}) \\
$^*$ From the SED of Fig.~\ref{sed} \\
\end{table*}

\section{Discussion}
\label{discu}

The availability of high and low angular resolution observations of
\I\ allows us to paint a detailed picture of the source, and discuss the
nature of the core \II .

\subsection{Virial equilibrium and density structure} 
\label{stabilita}

\begin{figure}
\centerline{\includegraphics[angle=-90,width=8.5cm]{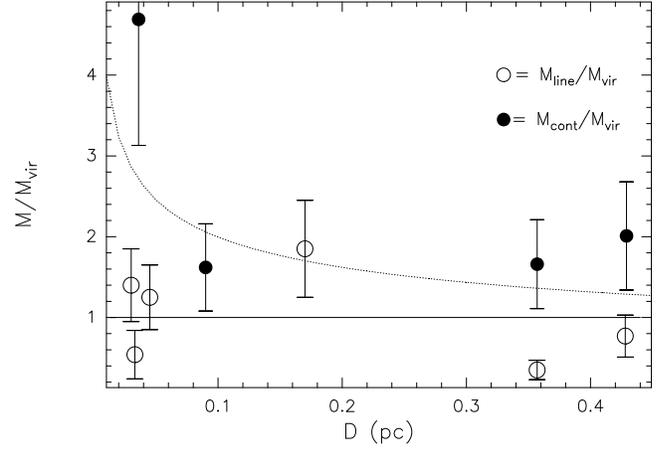}}
\caption{Ratio between masses deduced from gas ($M_{\rm line}$) or dust 
($M_{\rm cont}$) and the corresponding
virial mass as a function of the angular diameter. All values have been taken
 from Tables~\ref{tmass} and \ref{tmass2}. Filled circles assume
masses of the region estimated from dust, while empty circles assume
gas masses from the total column density of observed lines. The dotted line
represents the ratio $M/M_{\rm VIR}$, expected to be $\propto r^{-0.3}$
across the source. The estimated errors are of the order of $\sim30\%$.}
\label{mcd_mvir}
\end{figure}

The values of the masses estimated for the same region using dust and line 
emission show a systematic difference: from Table~\ref{tmass} 
and \ref{tmass2} one can see that the mass estimated from dust ($M_{\rm cont}$) is $\sim$ 2-3
times greater than that deduced from molecular
lines ($M_{\rm line}$). 
However, $M_{\rm line}$ is affected by several
problems which may lead one to underestimate the mass, as we will discuss
in Sect.~\ref{depletion}. On the other hand, the dust emission is independent 
of molecular abundances, so that the mass estimated 
from the mm continuum seems to be more reliable.
Since the source is a candidate massive protostar, it is useful to 
discuss its virial equilibrium. In order to do this, we need an estimate
of the virial mass. We have used the diameters deduced from each tracer 
to give an estimate of the virial mass. Then, we have compared these 
estimates to those deduced from the column density derived from the same tracer,
and with that obtained from the continuum emission arising from the same
region.
The result is shown in Fig.~\ref{mcd_mvir}. The empty circles indicate the
estimates derived from lines, while the full circles represent the continuum.

\begin{figure}
\centerline{\includegraphics[angle=-90,width=8.5cm]{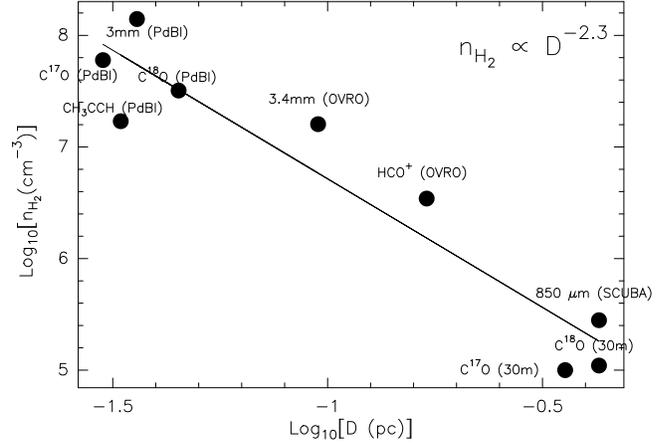}}
\caption{H$_2$ volume density against the linear diameter $D$ for all the 
tracers observed. All values have been taken from Table~\ref{tk}. The straight
line is a linear fit to the data, which gives 
${\rm n_{H_{2}}\propto D^{-2.3}}$.}
\label{nh2}
\end{figure}

In a cloud with a density profile which follows a power law of
the type $n(r)\propto r^{-m}$, one can demonstrate that the mass 
contained inside a given radius $r$, $M(r)$, is proportional to $r^{3-m}$. 
Instead, the virial mass depends on the source 
diameter and on the linewidth. However, based on the
values in Table~\ref{tmass}, we conclude that the linewidth does not change 
significantly at different radii. Therefore, the estimate of the ratio 
$M(r)/M_{\rm VIR}$ scales approximately as $r^{2-m}$. For \I , as
we shall show later (see Fig.~\ref{nh2}), we derive that
m=2.3$\pm$0.2, hence $M/M_{\rm VIR}\propto r^{-0.3}$. 
In Fig.~\ref{mcd_mvir} we plot the ratio $M/M_{\rm VIR}$ as a function of
$D$. The curve is in rough agreement with data points obtained from
the continuum measurements. However, the ``true'' virial mass, deduced
from the virial theorem, must be computed from the radius of the whole
cloud and the corresponding $\Delta v_{1/2}$. 
We do not know
the ``total'' dimension of the source, but we can compute at which radius 
the ratio $M(r)/M_{\rm VIR}$ becomes equal to 1.
For a power-law density distribution of the type 
$n(r)\propto r^{-m}$, the virial mass obtained from Eq.~(\ref{emvir}) must be 
multiplied by
 a factor $\frac{3}{5}\frac{5-2m}{3-m}$ (see MacLaren et 
al. \cite{mclaren}), which is $\sim0.35$ for m=2.3. This
increases the ratio $M/M_{\rm VIR}$ by a factor 3, implying that 
$M/M_{\rm VIR}\geq 3$ at a radius of 10\asec , and that the
virial mass
equals the gas mass at a radius of $\sim 50$ \asec , which is much greater than
the maximum diameter observed by us. Hence, since the real source size
is likely to be smaller than that corresponding to $M/M_{\rm VIR}=1$, we 
believe that \II\ is likely to be gravitationally unstable.
This conclusion is further supported by discussing the density profile.

In Fig.~\ref{nh2} we plot the density estimated from all tracers observed
against the corresponding linear diameter (densities
and diameters are given in Table~\ref{tmass} and~\ref{tmass2}). The data 
follow the relation $n_{\rm H_{2}}\propto D^{-2.3}$, hence 
$n_{\rm H_{2}}\propto r^{-2.3}$ . This
is similar to that predicted by star formation models (Shu et al.
\cite{shu}), in which star formation occurs in molecular
clouds which are singular isothermal spheres with
density $\propto r^{-2}$. This sphere rapidly undergoes inside-out
collapse.
Then, the density profile in Fig.~\ref{nh2} supports the idea that \I\ is 
unstable, as expected for a protostar which is still in the accretion phase.
 
\subsection{A depleted core?}
\label{depletion}

On the basis of Tables~4-6 and the SCUBA map of Fig.~\ref{scuba_co}, 
one can identify two regions in \II :
a compact core and a surrounding extended halo. Their physical parameters 
are:
\begin{itemize}
\item $D_{1}=0.035$ pc, $T_{1}\simeq42$ K, $n_{1}\simeq 14.0\;10^{7}$ \cmc\
for the core;
\item $D_{2}\simeq0.4$ pc, $T_{2}\simeq15$ K, $n_{2}\simeq 3\;10^{5}$ \cmc\ 
for the halo. 
\end{itemize}
where the diameters and $n_{\rm H_{2}}$ volume densities have been derived
from the continuum measurements, and the kinetic temperatures from the lines.
Now, using a simple source model, we want to understand if it is possible 
to clarify 
the absence of a central peak in single-dish maps, and also the 
non-detection of \CIII\ (1--0) in the PdBI observations. 

\begin{figure}[!h]
\centerline{\includegraphics[angle=-90,width=8.0cm]{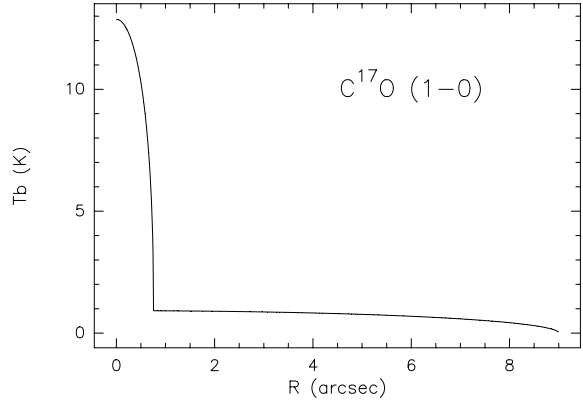}}
\caption{Brightness temperature of the
\CIII\ (1--0) line computed assuming a source as described in the text, and
a \CIII\ relative abundance of $1.4\;10^{-8}$.
$R$ represents the radial distance from the source center.}
\label{shell}
\end{figure}

\begin{figure}[!h]
\centerline{\includegraphics[angle=-90,width=7.5cm]{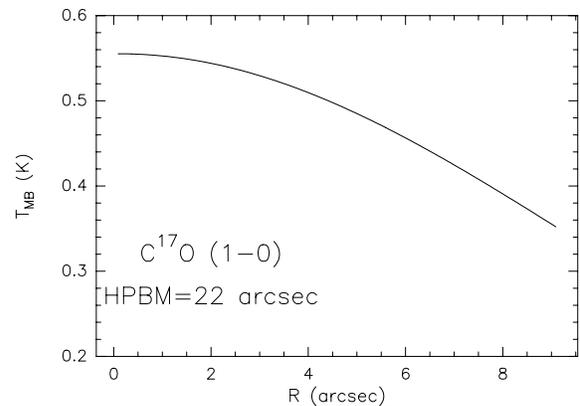}}
\caption{Brightness temperature of Fig.\ref{shell} convolved
with the IRAM-30m telescope beam (22\asec).}
\label{tmb_42k}
\end{figure}

Let us consider a source consisting of the two regions outlined above:
a core and a surrounding halo. Both are assumed to be spherical, homogeneous
and isothermal. 
The expected brightness temperature $T_{\rm B}$ of the \CIII\ (1--0) line for
such a source is shown in Fig.~\ref{shell}, where 
$T_{\rm B}$ is plotted against the radial distance $R$ from the source center.
Convolving $T_{\rm B}$ with a gaussian beam, we derive the 
corresponding main beam brightness temperature
 $T_{\rm MB}$. Fig.~\ref{tmb_42k} represents $T_{\rm B}$ convolved with a 
gaussian beam with HPBW=22\asec\ (that of the IRAM 30-m telescope); one 
can see that the emission coming from the core is flattened by beam dilution 
effects. 

\begin{figure*}[h*]
\centerline{\includegraphics[angle=-90,width=14cm]{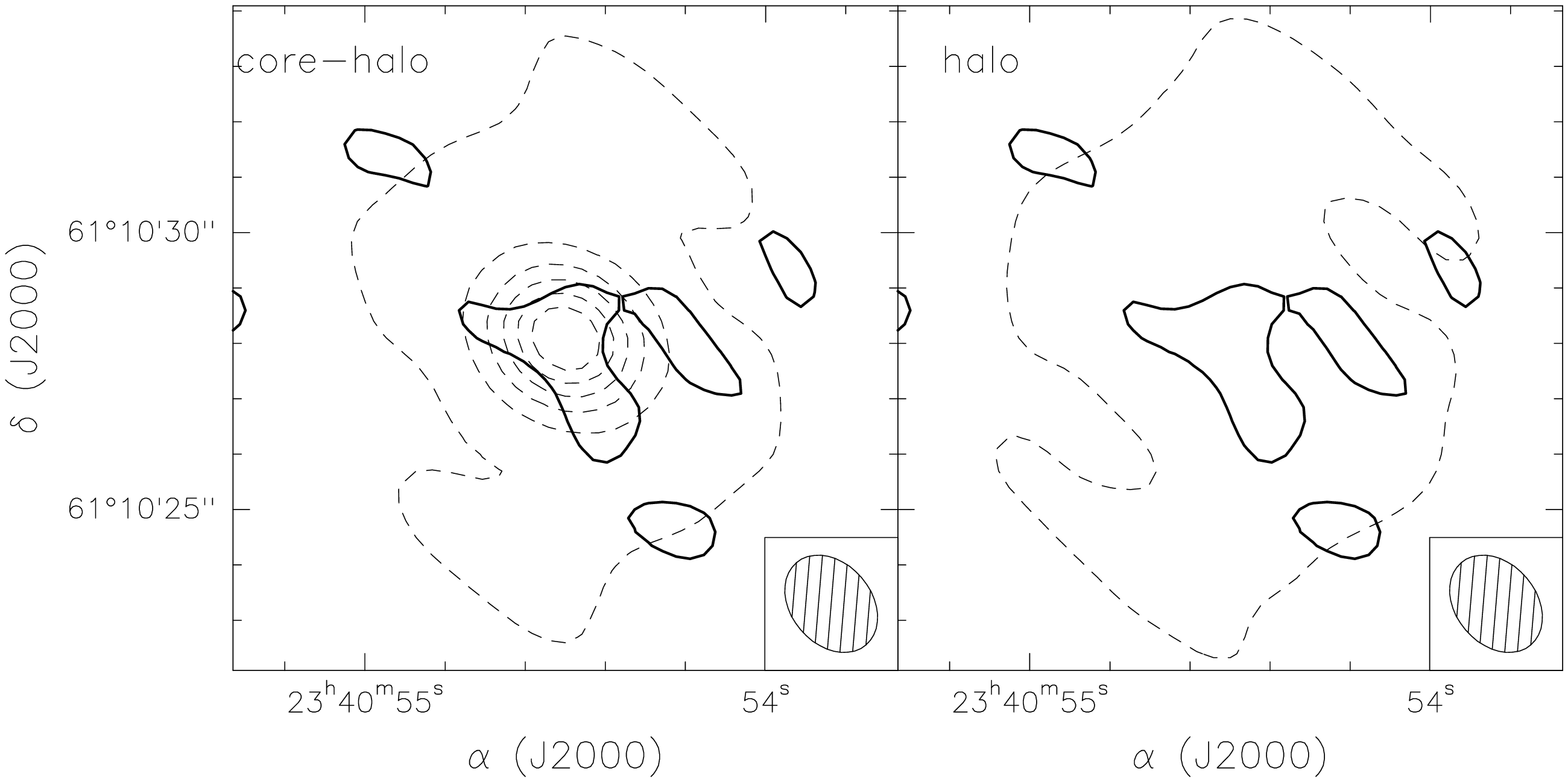}}
\caption{Left panel: map of the \CIII\ (1--0) line in the peak channel 
(at $-50.5$ \kms\ ). The solid contour represents the observed 2$\sigma$ level 
($\sim 0.024$ Jy beam$^{-1}$), while the dashed contours indicate the emission
of a ``core-halo'' source model, as described in Fig.~\ref{shell}. Levels range
from $0.024$ to 0.14 by 0.024 Jy beam$^{-1}$. Right
panel: same as left for a source model without the central core. Both
contours represent the $2\sigma$ observed level.}
\label{shell_model}
\end{figure*}

We now discuss why the central core is not
detected in the \CIII\ (1--0) PdBI map. Since the molecular emission is extended, we
must determine first how much flux density is filtered out by the
interferometer. Starting from the 
brightness temperature profile of Fig.~\ref{shell}, we have used a 
modified version of the procedure described in Wilner et al. 
(\cite{wilner2}) and Testi et al. (\cite{testi}). The program uses
the model to create first the source image in the sky; then, it
computes the Fourier Transform and resamples the visibility function in the 
UV points corresponding to those measured in that channel, which finally are 
used to reproduce the image.
In order to compare the PdBI map with what is predicted by the model, we plot 
in Fig.~\ref{shell_model} the channel where the peak is observed 
(at $-50.5$ \kms ), superimposed on the map predicted by our model. 
We have also done the same using a model that reproduces the ``halo'' only.
It is evident that the model reproduces a central peak which is 
instead completely lost in the PdBI map; on the other hand,
the emission of the halo is quite consistent with what observed if we 
consider that our model is very simple, and does not consider possible 
patchy structures inside the halo. 

A possible explanation of the discrepancy between the PdBI map and our model
 can be that in the core the
\CII\ and \CIII\ species are partially depleted because they are frozen
onto dust grains. This is suggested by the fact that gas
masses deduced from the CO isotopomers given in Table~\ref{tmass} are 
systematically
lower than the dust masses inferred from the continuum measurements. This 
might be due to a partial freeze-out of the molecules on the dust grains. 
This situation is possible even at temperatures as high as
40 K if dust grains have ice mantles. In fact, chemical models predict that
ice mantles may exist at the densities and 
temperatures measured by us in the core (Van Dishoeck \& Blake 
\cite{vandish}), and this allows partial depletion of CO and its isotopomers.
Furthermore, recently Thompson \& Macdonald (\cite{thompson}) have 
performed a chemical analysis of \II\ using molecular lines with rest
frequencies in the 
range 330-360 GHz, finding that the chemical composition of \II\ is 
consistent with that of a 
molecular core in the middle evaporation phase, i.e. when the majority of the
molecular species are beginning to be evaporated from dust grains ice
mantles.

Also, dust grains with ice mantles
might be responsible for the non-detection of NH$_3$ towards the source center:
in fact, the same chemical models also predict that the presence of ice 
mantles allows NH$_3$ depletion up to temperatures of 90 K.
Finally, it is worth noting that partial depletion of CO is consistent
 with detection of \ace\ emission in the core. In fact, Ruffle et al. 
(\cite{ruffle}) found that in dense cores, when CO depletion occurs, 
molecular species produced from CH and without oxygen (like \ace ) increase
their rate of production.

\subsection{The nature of \II }
\label{discu_core}

Here we want to find out whether our results may be used to establish the
nature and the evolutionary stage of the embedded YSO.

\begin{figure}
\centerline{\includegraphics[angle=-90,width=8.0cm]{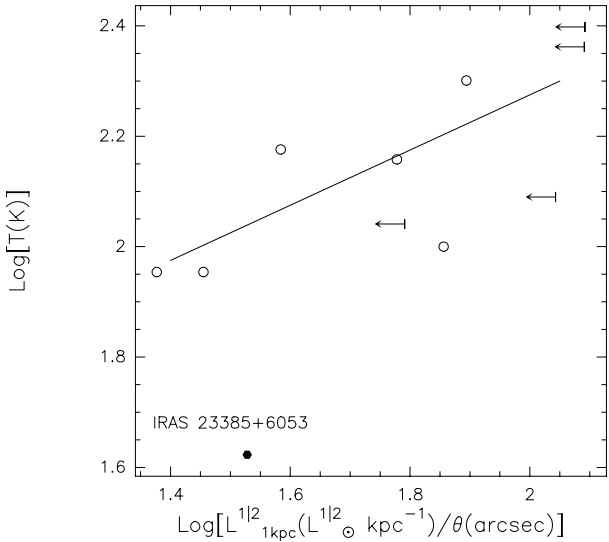}}
\caption{Kinetic temperature measured at the surface of molecular cores 
plotted against the distance-independent ratio $L^{1/2}/\theta$, where $\theta$
is the angular diameter of the core, and $L$ the luminosity of the embedded
(proto)star. Open circles indicate HCs (Kurtz et al. \cite{kurtz})
while the filled circle corresponds to 
\II . The straight line represents the steepest power law 
for spherical cores, $T\propto(L^{1/2}/\theta d)^{0.5}$,
expected for a dusty core heated by an embedded star 
(Doty \& Leung~\cite{doty}).}
\label{hotcores}
\end{figure}

In cores where the gas is heated by an embedded star, the gas temperature
at a radial distance $R=\theta d$ from the central star is expected to
scale as (see e.g. Plume et al. \cite{plume}):
\begin{equation}
T\propto\left(\frac{L^{1/2}}{\theta d}\right)^{\alpha}
\label{etemp}
\end{equation}
where $L$ is the luminosity of the embedded star. Eq.~(\ref{etemp}) represents
the gas temperature, but at densities as high as $10^{6}$ \cmc\ we can
assume coupling between gas and dust. Hence, at such densities 
Eq.~(\ref{etemp}) also describes
the dust temperature.
The index $\alpha$ depends on the opacity index, $\beta$, (Doty \& Leung
\cite{doty}), and it lies between 0.3 and 0.5. The steepest power law is 
obtained when $\alpha=0.5$, and it corresponds to the 
optically thick case, when the luminosity irradiated at the core surface
$R$ equals the stellar luminosity, namely:
\begin{equation}
L_{*}=4\pi R^{2}\sigma T^{4}\;,
\label{elum}
\end{equation}
where $T$ represents the surface temperature.
If instead the core is optically thin, the temperature profile is flatter.
Therefore, the core temperature predicted by Eq.~(\ref{elum}) represents a
``lower limit'', and at 0.017 pc (corresponding to the radius of \II\ deduced
from the \ace\ PdBI emission) for
a central star with luminosity $L\sim 1.6\;10^{4}L_{\odot}$ this lower
limit is $\sim 51$ K. The temperature derived from \ace\ is significantly
lower than that. This discrepancy is confirmed by theoretical models: Osorio \&
Lizano (\cite{osorio}) have performed a theoretical simulation of the
temperature profile of
\II\ assuming a source luminosity of $\sim 10^{4}L_{\odot}$, deriving a
temperature much higher than 42 K at the same radius (see their 
Fig.~5). To compare our result with what found in HCs, in 
Fig.~\ref{hotcores} we plot the temperature $T$ at the core surface
versus $L^{1/2}/R$, for a
sample of HCs and for \II . The luminosity $L$ of the central
star is given by the IRAS luminosity for HCs without an embedded \HII\ 
regions, and it represents an upper limit. For those with 
an embedded \HII\ region, the luminosity of the central star has been
deduced from the radio continuum flux of the embedded \HII\ region 
(Wood \& Churchwell \cite{w&c}).

The drawn line indicates the steepest power
law, $T\propto(L^{1/2}/R)^{0.5}$. It is evident that it is not possible
to match our temperature estimate with that expected for such a power law. 
To solve this problem we examine three possibilities:

1) We might be overestimating the source luminosity.
In Sect.~\ref{contspec} we discussed that
the luminosity of the embedded source is of the order of $10^{4}$ solar
luminosities. This was based on the hypothesis that the bulk of the IRAS
fluxes at 60 and 100 $\mu$m is due to \II . Unfortunately, the IRAS angular
resolution is not sufficient to confirm this hypothesis.
Hence, the ``bright ring'' surrounding \II\ might emit a consistent
fraction of the flux at 60 and 100 $\mu$m. In that case, the luminosity of \II\ 
might be lower. A temperature of 42 K at 0.017 pc in an optically thick core 
implies a central source of $\sim 1.5\;10^{3}L_{\odot}$. Altough well below 
the value predicted in Sect.~\ref{micont}, this luminosity is 
consistent with an intermediate-mass protostar deriving its luminosity from 
accretion, as shown by Behrend \& Maeder (\cite{behrend}). 
In fact, the authors performed 
calculations of the evolution of protostars with mass from 1 to 85 
$M_{\odot}$, assuming growing accretion rates. For an accretion luminosity of 
$\sim 1.5\;10^{3}L_{\odot}$, the protostar mass is 
$\sim 6\;M_{\odot}$.

2) Another possibility is that we have underestimated the core diameter.
This is possible if the 
source is not gaussian: in fact, if the core is spherical, the diameter
estimated in Table~\ref{tmass} requires a correction (see Panagia \& Walmsley 
\cite{pana}). In our case one obtains $\sim 2$\asec (0.047 pc), for the 
diameter of the core. This brings the luminosity
to $3\;10^{3}{\rm L_{\odot}}$, still much less than the value obtained in 
Sect.~\ref{contspec} from the continuum spectrum.

3) Finally, it is possible that the temperature profile,
is steeper in case of a non-spherical geometry, as in the presence of
a disk. Cesaroni et al. (\cite{cesa98}) have found that HCs are likely to
be disk-like structures, in which the temperature profile is 
$T\propto R^{-0.75}$.
With this power-law it is possible to obtain a lower temperature 
at 0.017 pc with a luminosity of 1.6 $10^{4} {\rm L_{\odot}}$.

Which solution is correct? Concerning the second hypothesis
(underestimate of the core radius),
we have already seen that it is very unlikely.
This is also valid for the third one (disk-like structures): in fact, to 
measure such a low temperature, the disk should 
lie along the line of sight, but unfortunately this is more likely
``face-on'', because the outflow detected by Molinari et al. (\cite{mol98b})
lie approximately along the line of sight. The first hypothesis 
(overestimate of the core luminosity) is the most likely, but
presently it is impossible to give a reliable answer,
because the correct source luminosity can be derived only by means of a map 
at $\sim$ 100 $\mu$m at high angular resolution, which is not available. 
However, as already discussed, a luminosity
as low as $10^{3}$ $L_{\odot}$ is consistent with a protostar in the accretion 
phase, whose mass, at present, is $\sim6\;M_{\odot}$. 

\section{Conclusions}

We have used the IRAM-30m telescope, the Plateau de Bure Interferometer and 
the Very Large Array to observe the massive protostar candidate \I\ in
the \CII\ (1--0), \CIII\ (1--0) and (2--1), \ace\ (6--5), (8--7) and (13--12),
\AMM\ (1,1) and (2,2) lines. The following results have been obtained:
\begin{itemize}
\item The source consists of two regions: a compact molecular core 
with a diameter $\sim 0.03\div 0.04$ pc and a halo which has a diameter of
up to $\sim 0.4$ pc; the core
has a kinetic temperature of $T_{\rm k}\sim 42$ K and a $\rm H_{2}$ volume 
density of the order of $n_{\rm H_{2}}\sim 10^{7}$ \cmc ; the halo has an 
average temperature $T_{\rm k}\sim 15$ K and an average $\rm H_{2}$ volume 
density $n_{\rm H_{2}}\sim 3\; 10^{5}$ \cmc ;
\item based on the continuum spectrum, the source luminosity can vary
between $1.6\;10^{4}$ and $1.5\;10^{2}\;L_{\odot}$: we believe
that the upper limit is closer to the real source luminosity; 
\item by comparing the $\rm H_{2}$ volume density derived at different
radii from the IRAS source, we deduce that the halo has a density profile
of the type $n_{\rm H_{2}}\propto r^{-2.3}$, suggesting that it 
might be gravitationally unstable. This is further supported by the fact
that, assuming a density profile as above, the virial mass is much lower
than the gas mass. 
\item On the basis of the steepest theoretical temperature profile 
expected
in cores heated by embedded stars ($T\propto R^{-0.5}$), it is impossible 
to explain a temperature of 42 K at the core radius, if the source 
luminosity is $L\sim 10^{4}{\rm 
L_{\odot}}$. We propose three possible solutions: 1) we are overestimating
the stellar luminosity. A temperature of $42$ K at the core radius ($\sim
0.017$ pc) is possible if the stellar luminosity is 
$\sim 10^{3}{\rm L_{\odot}}$. This solution is consistent with 
an intermediate-mass protostar deriving its luminosity from accretion; 
2) we are underestimating the core radius in case of non-gaussian source, 
but this solution is unlikely because the correction that we should apply is 
very small; 3) the source is not spherical: in that case
the temperature profile is expected to be steeper ($T\propto R^{-0.75}$) and
it is possible to reach a temperature as low as 42 K at the core radius even
with luminosities as high as $10^{4}{\rm L_{\odot}}$.

The first hypothesis seems the most likely, although the real source 
luminosity can be derived only with
high resolution maps at FIR wavelengths, which will become feasible only
with the HERSCHEL satellite.
\end{itemize}

\begin{acknowledgements}
We would like to thank the referee, Dr. H. Beuther, for his very useful
comments. We also thank Paola Caselli for stimulating discussion about
the main topic of Sect.~5.2.  
\end{acknowledgements}
{}

\end{document}